\documentstyle[psfig,12pt,citesort]{article}

\def \el {{\ell}}

\def \ov {\over}

 \def\ep {\epsilon}

\def \t {\theta}
\def \p {\phi}
\def \ep {\epsilon}

\def \ps {\psi}

\newcounter{subequation}[equation]

\newcommand{\be}{\begin{equation}}
\newcommand{\ee}{\end{equation}}
\newcommand{\eel}[1]{\label{#1}\end{equation}}
\newcommand{\bea}{\begin{eqnarray}}

\newcommand{\eea}{\end{eqnarray}} \newcommand{\eeal}[1]{\label{#1}\end{eqnarray}}

\makeatletter

\def\thesubequation{\theequation\@alph\c@subequation}
\def\@subeqnnum{{\rm (\thesubequation)}}
\def\slabel#1{\@bsphack\if@filesw {\let\thepage\relax
   \xdef\@gtempa{\write\@auxout{\string
      \newlabel{#1}{{\thesubequation}{\thepage}}}}}\@gtempa
   \if@nobreak \ifvmode\nobreak\fi\fi\fi\@esphack}
\def\subeqnarray{\stepcounter{equation}
\let\@currentlabel=\theequation\global\c@subequation\@ne
\global\@eqnswtrue
\global\@eqcnt\z@\tabskip\@centering\let\\=\@subeqncr

$$\halign to \displaywidth\bgroup\@eqnsel\hskip\@centering
  $\displaystyle\tabskip\z@{##}$&\global\@eqcnt\@ne
  \hskip 2\arraycolsep \hfil${##}$\hfil
  &\global\@eqcnt\tw@ \hskip 2\arraycolsep
  $\displaystyle\tabskip\z@{##}$\hfil
   \tabskip\@centering&\llap{##}\tabskip\z@\cr}
\def\endsubeqnarray{\@@subeqncr\egroup
                     $$\global\@ignoretrue}
\def\@subeqncr{{\ifnum0=`}\fi\@ifstar{\global\@eqpen\@M
    \@ysubeqncr}{\global\@eqpen\interdisplaylinepenalty \@ysubeqncr}}
\def\@ysubeqncr{\@ifnextchar [{\@xsubeqncr}{\@xsubeqncr[\z@]}}
\def\@xsubeqncr[#1]{\ifnum0=`{\fi}\@@subeqncr
   \noalign{\penalty\@eqpen\vskip\jot\vskip #1\relax}}
\def\@@subeqncr{\let\@tempa\relax
    \ifcase\@eqcnt \def\@tempa{& & &}\or \def\@tempa{& &}
      \else \def\@tempa{&}\fi
     \@tempa \if@eqnsw\@subeqnnum\refstepcounter{subequation}\fi
     \global\@eqnswtrue\global\@eqcnt\z@\cr}
\let\@ssubeqncr=\@subeqncr
\@namedef{subeqnarray*}{\def\@subeqncr{\nonumber\@ssubeqncr}\subeqnarray}

\@namedef{endsubeqnarray*}{\global\advance\c@equation\m@ne%
                           \nonumber\endsubeqnarray}

\makeatletter \@addtoreset{equation}{section} \makeatother
\renewcommand{\theequation}{\thesection.\arabic{equation}}

\catcode`\@=11

\newcount\hour
\newcount\minute
\newtoks\amorpm \hour=\time\divide\hour by 60\minute=\time{\multiply\hour by 60 \global\advance\minute by-\hour}
\edef\standardtime{{\ifnum\hour<12 \global\amorpm={am}%
        \else\global\amorpm={pm}\advance\hour by-12 \fi
        \ifnum\hour=0 \hour=12 \fi
        \number\hour:\ifnum\minute<10
        0\fi\number\minute\the\amorpm}}
\edef\militarytime{\number\hour:\ifnum\minute<10
0\fi\number\minute}

\def\draftlabel#1{{\@bsphack\if@filesw {\let\thepage\relax
   \xdef\@gtempa{\write\@auxout{\string
      \newlabel{#1}{{\@currentlabel}{\thepage}}}}}\@gtempa
   \if@nobreak \ifvmode\nobreak\fi\fi\fi\@esphack}
        \gdef\@eqnlabel{#1}}
\def\@eqnlabel{}
\def\@vacuum{}
\def\marginnote#1{}
\def\draftmarginnote#1{\marginpar{\raggedright\scriptsize\tt#1}}
\def\draft{
        \pagestyle{plain}
        \overfullrule=2pt
        \oddsidemargin -.5truein
        \def\@oddhead{\sl \phantom{\today\quad\militarytime} \hfil
        \smash{\Large\sl DRAFT} \hfil \today\quad\militarytime}
        \let\@evenhead\@oddhead
        \let\label=\draftlabel
        \let\marginnote=\draftmarginnote
        \def\ps@empty{\let\@mkboth\@gobbletwo
        \def\@oddfoot{\hfil \smash{\Large\sl DRAFT} \hfil}
        \let\@evenfoot\@oddhead}

\def\@eqnnum{(\theequation)\rlap{\kern\marginparsep\tt\@eqnlabel}%
        \global\let\@eqnlabel\@vacuum}  }

\renewcommand{\theequation}{\thesection.\arabic{equation}}
\renewcommand{\thefootnote}{\fnsymbol{footnote}}

\def \ov {\over}

 \def\ep {\epsilon}

\def \t {\theta}
\def \p {\phi}
\def \ep {\epsilon}

\def \ps {\psi}

\def\o{\omega}

\def\pd{\partial}

\def\m{\mu}

\def\a{\alpha}
\def\b{\beta}

\def\s{\sigma}

 \def \t {\theta}

\def\p{\phi}

\def\ep{\epsilon}

\def \foot{ \footnote}
\def\be{\begin{equation}}
\def\ee{\end{equation}}

\def \m {\mu}

\newcommand{\baq}{\begin{equation}\begin{array}{rcl}}
\newcommand{\eaq}{\end{array}\end{equation}}
\newcommand{\eaql}[1]{\end{array}\label{#1}\end{equation}}
\newcommand{\beac}{\begin{equation}\begin{array}{rcl}}
\newcommand{\eeacn}[1]{\end{array}\label{#1}\end{equation}}
\newcommand{\ba}{\begin{array}}
\newcommand{\ea}{\end{array}}

\def\rem#1{}

\def\none{${\cal N}=1$}
\def\ntwo{${\cal N}=2$}
\def\ZZ{{\bf Z}}
\newcommand\ket[1]{|#1\rangle}
\newcommand\eref[1]{(\ref{#1})}

\begin{document}

\begin{titlepage}

\hfill hep-th/0212061

\hfill MCTP-02-45

\hfill UW/PT 02-28

\hfill CERN-TH/02-211

\begin{center}
\vskip 0.5 cm

{\Large \bf A Soluble String Theory of Hadrons}%

\vskip .7 cm

\vskip 1 cm

{\large E.G. Gimon${}^1$,  L.A. Pando Zayas${}^{1,2}$,
J. Sonnenschein${}^{1,3,4}$ and  M.J. Strassler${}^{1,5}$}\\

\end{center}

\vskip .4cm
\centerline{\it ${}^1$ School of Natural Sciences, Institute for Advanced Study}
\centerline{\it  Princeton, NJ 08540}

\vskip .4cm \centerline{\it ${}^2$ Michigan Center for Theoretical
Physics}
\centerline{ \it Randall Laboratory of Physics, The University of
Michigan}
\centerline{\it Ann Arbor, MI 48109-1120}

\vskip .4cm \centerline{\it ${}^3$ School of Physics and Astronomy}
\centerline{ \it Beverly and Raymond Sackler Faculty of Exact Sciences}
\centerline{ \it Tel Aviv University, Ramat Aviv, 69978, Israel}

\vskip .4cm \centerline{\it ${}^4$ Theory Division, CERN}
\centerline{ \it CH-1211 Geneva 23, Switzerland}

\vskip .4cm \centerline{\it ${}^5$ Department of Physics and Astronomy}
\centerline{ \it P.O Box 351560, University of Washington}
\centerline{\it Seattle, WA 98195}

\vskip 1.5 cm

\begin{abstract}
We consider Penrose limits of the Klebanov-Strassler and
Maldacena-N\'u\~nez holographic duals to ${\cal N}=1$
supersymmetric Yang-Mills. By focusing in on the IR region we
obtain exactly solvable string theory models. These represent the
nonrelativistic motion and low-lying excitations of heavy hadrons
with mass proportional to a large global charge. We argue that
these hadrons, both physically and mathematically, take the form
of heavy nonrelativistic strings; we term them ``annulons.'' A
simple toy model of a string boosted along a compact circle allows
us considerable insight into their properties. We also calculate
the Wilson loop carrying large global charge and show the effect
of confinement is quadratic, not linear, in the string tension.

\end{abstract}

\end{titlepage}

\setcounter{page}{1}
\renewcommand{\thefootnote}{\arabic{footnote}}
\setcounter{footnote}{0}

\def \N{{\cal N}} \def \ov {\over}

\section{Introduction and Summary}

Since the experimental observation of stringy behavior in hadronic
physics, it has been hoped that these aspects of the strong
interactions could be predicted from an as-yet-unknown string theory.
The recent discovery of a precise duality of gauge theory and string
theory has allowed some progress.  The string theoretic descriptions
of several confining gauge theories with a large number of colors have
now been found.  More precisely, the backgrounds on which the dual
type IIB strings propagate are now known.  However, the string theory
on these backgrounds is not soluble.  This is unfortunate: while the
masses of low-lying low-spin hadrons can be computed from supergravity
(SUGRA), the full hadron spectrum requires string theory.  SUGRA
cannot describe states of high spin and cannot see the Regge
trajectories that we would expect gauge theory to exhibit.

 Initially most information gained from gauge/string duality
was obtained through analysis of the SUGRA backgrounds, or of
semiclassical branes and/or strings in these backgrounds.  However,
this has begun to change \cite{jpms,bmn,gkp}. The remarkable work of
Berenstein, Maldacena and Nastase (BMN) \cite{bmn} has provided a link
between an exactly solvable worldsheet string theory \cite{metsaev}
and a sector of the conformal ${\cal N}=4$ supersymmetric Yang-Mills
(SYM) theory.  This work was extended to other conformal theories in
\cite{ikm,go,PS}.  Attempts to apply the Penrose limit
\cite{penrose,gueven,blau} to non--conformal backgrounds
\cite{PS,GPS,cvj} have resulted in string theories with world-sheet
time-dependent mass terms. The geodesics chosen in these cases are
appropriate to the study of the properties of the
renormalization-group (RG) flow. The corresponding world-line problem
in the Pilch-Warner background solution is exactly solvable
\cite{GPS}, giving the ``branching'' of a given operator in the
ultraviolet (UV) ${\cal N}=4$ SYM into operators of the infrared (IR)
\none\ theory (the conformal theory with two adjoint chiral multiplets.)

In this paper we study theories that exhibit confinement and a
discrete spectrum of hadrons in the IR.  Our interest is not in
operators but in hadronic states of fixed mass in Minkowski space;
consequently we choose a different type of geodesic as a basis
for our Penrose limit.  By focusing on geodesics that are frozen
at the minimal $AdS$ radius in the IR, and that spiral inside the
cylinder formed by the Minkowski time direction and a circle in
the compact part of the ten dimensional space, we can obtain an
exactly solvable time--independent string theory background in the
Penrose limit which captures the dynamics of hadrons with a large
global charge.

  The specific confining gauge theories we consider consist of \none\
SYM plus massive particles in the adjoint
representation and carrying a global abelian charge.  The string
Hamiltonian describes hadrons which are bound states of these massive
particles, in the limit that the global charge and the number of
colors both go to infinity.  Roughly speaking, the string sigma model
takes the form of a ten-dimensional string, which in light-cone gauge
is ``compactified'' by world-sheet mass terms down to the three
massless spatial dimensions of Minkowski space.  The vacuum of the
string theory is a stationary hadron of large mass and charge. Our
string Hamiltonian describes its non-relativistic motion (and that
of its fermionic superpartners) in three
spatial dimensions, and its low-lying stringy excitations in those
directions, as well as
excitations which add a small number of other globally-charged
constituents.  We argue that these hadrons take the physical form of
non--relativistic strings.

In backgrounds corresponding to confining gauge theories, there
is a minimum $AdS$ radius $r_0$, where $g_{tt}$ generally goes to
a non-zero minimum.  In section 2 we use this, along with mild
constraints on the space perpendicular to the branes, to find
null geodesics fixed at $r_0$ (other ``frozen'' geodesics appear
in \cite{frozen}) . We explicitly discuss the two trademark SUGRA
solutions dual to ${\cal N}=1$ SYM in the IR: the
Maldacena-N\'u\~nez (MN) background \cite{MN}, in Section 3, and
the Klebanov-Strassler (KS) \cite{ks,kh} background (with a
nonstandard but convenient parameterization), in Section 4.
Section 5 describes the light-cone quantization and spectrum of
the corresponding string theories, with comments about the
unbroken supersymmetries. In Section 6 we find a hadronic
interpretation of the string spectrum for the KS case. (The MN
case is similar but less well understood.)  In particular, we show
that a very simple toy model (a string moving on a compact circle)
captures some of the terms of the string Hamiltonian, thereby
emphasizing its universality.

 In section 7 we obtain, under very general assumptions, an
expression for the Wilson loop with global charge. We arrive at the same
formula for the Wilson loop using heuristic
field theory arguments, the above-mentioned toy model,  and
a semi--classical string analysis.   We close with a few
comments and include three appendices.   Appendix A contains an
explicit derivation of the new parameterization of the deformed
conifold and its relation to the standard coordinates. In
appendix B we present the main steps in the derivation of the
string Hamiltonian. Appendix C contains some technical arguments
about the reliability of the Wilson loop ansatz used in section 6.

\section{Null geodesics in the IR of confining theories}

  Let us first clarify why we choose to study null geodesics at the
minimal $AdS$ radius.  The essential feature of a confining theory is
that it has stable electric flux tubes and a spectrum of hadrons of
definite (four-dimensional) mass. A hadron of definite
four-dimensional mass is a supergravity eigenstate of the
ten-dimensional Laplacian which is also an eigenstate of the
four-dimensional Minkowski Laplacian.  (This is in contrast to an
operator of definite dimension, which is an eigenstate of the
five-dimensional $AdS$ Laplacian.)  These states are plane waves in
the Minkowski directions and have nontrivial wave functions
$\psi(r,\Omega)$ on the remaining six directions; here $r$ is the
$AdS$ radius (which extends from the boundary at $r\to\infty$ to a
finite minimum at $r=r_0>0$.)  A hadron's wave function falls off as
$r^{-\Delta}$, where $\Delta$ is the dimension of the lowest-dimension
operator which can create the hadron.  A hadron of large charge $J$
--- which is typically heavy, $m\sim J$, since it has many
constituents of charge 1 --- can be created only by an operator of
large charge, which, since it contains of order $J$ fields, has
$\Delta\sim J$.  Consequently {\em hadrons of high charge correspond
to modes which are concentrated close to $r=r_{0}$}.\footnote{This is
one of many examples which caution that one must avoid naive
application of the dictum that $AdS$ radius is the same as energy;
baryons represent another.}

Since these phenomena are localized at $r=r_{0}$, we should expect
they are sensitive mainly to the IR physics of the gauge theory,
and should not depend much on the UV completion of the low-energy
theory. For this reason, we expect only mild differences between
the MN and KS examples, which are both ${\cal N}=1$ SYM in the
IR; although their geometries differ greatly in the UV, they are
similar in the IR and we would expect the plane wave string
theories are also similar.  In either background, it is natural
to look for geodesics with $r=r_0$ and $\dot r=0$, $\dot t = 1$,
where a dot represents a derivative with respect to the affine
parameter of the geodesic.
With this choice, the pp-wave Hamiltonian will
measure not dimensions of operators (alternatively, energies of
states on a spatial $S^3$) but rather energies of states in
Minkowski space (with a flat spatial ${\bf R}^3$.)

 In order for a geodesic at a fixed radius to be null (the key
ingredient for a consistent Penrose limit), it must move both in time
and in a bulk spatial direction.  Requiring isotropy in the three
spatial dimensions of the gauge theory, along with $\dot r=0$, forces
us to choose the geodesic to move on a curve inside the other bulk
dimensions, typically a closed circle generated by a Killing
vector.  Consequently the states in the dual gauge theory will
carry large charge under the corresponding global symmetry.
Their spins, by contrast, will be of order one.

The  conditions for a null geodesic of this type are easily found.
The time $t$  and the radial direction $r$ are  effectively described
by the following Lagrangian
\be
 \label{simple}
 {\cal L}=-g_{tt}\dot{t}^2 + g_{rr}\dot{r}^2 +
g_{\p\p}\dot{\p}^2,
\ee
where dot means differentiation with respect to the affine
parameter $u$.  Assuming for simplicity that the metric depends
only on the  radial coordinate (which is approximately true in
some neighborhood of interest), the equations of motions are
\bea
\label{eom}
\dot t &=& {E\over g_{tt}}, \qquad  \dot \p= {\mu\over
g_{\phi\phi}}, \nonumber \\ 2{d\over
du}\left(g_{rr}\dot{r}\right)&=& \dot{r}^2\pd_r\,
g_{rr}-\dot{t}^2 \pd_r \, g_{tt}  +\dot{\p}^2 \pd_r \, g_{\p\p} \ .
\eea
There is also a constraint, ${\cal L}= 0$,  which we re--write
using the  equations of motion  for $t$ and $\phi$
\be
\label{constraint}
g_{rr}\,\dot{r}^2 = {E^2\over g_{tt}}- {\mu^2\over g_{\p\p}}.
\ee
Since we are interested in geodesics at a fixed radius $r=r_0$ we
impose the condition $\dot{r}|_{r_0}=0$. Thus (\ref{constraint})
tells us that in the neighborhood of $r_0$ we have
\be
\label{cond1}
{E^2\over g_{tt}} = { \m^2\over g_{\p\p}}.
\ee
In the case of confining theories we have $g_{tt}(r_0)> 0$, so we
can satisfy this equation for any finite $g_{\p\p}(r_0)$ by
adjusting $\m$.

Since the geodesic equation is second order, the acceleration,
$\ddot{r}$, must also vanish for our geodesic to remain at a
fixed value of $r$. Looking at the equations of motion
(\ref{eom}) we see that such a condition implies
\be
\label{cond2}
\pd_r\,(g_{tt})= {E^2\over\mu^2}\,\pd_r\,g_{\p\p}.
\ee
For confining theories, there is a minimum $r =r_0$ for which
$\pd_r g_{tt}(r_0) = 0$.  If we assume that $g_{\p\p}$ depends
smoothly on $r$, and only through the magnitude $|r-r_0|$,
then Eq.~\eref{cond2} is also easily satisfied.

 It is interesting to note that the conditions (\ref{cond1}) and
(\ref{cond2}) derived here look very similar to the ones that
appear to describe confining theories in \cite{cobiwl} (derived
from considering Wilson loops) except that the metric component
$g_{\p\p}$ now plays an important role. This role comes from
looking at charged Wilson loops, as we shall see in section 7.

  To summarize, we are interested in looking at objects of
large charge in confining theories.  We know of several such
confining theories which have dual SUGRA descriptions when
embedded in useful UV theories.  It becomes easy to characterize
the objects we want, regardless of the vagaries of their UV
completion, if we look at the states localized near a null
geodesic at the minimum radius (confinement scale).  As we have
shown, such null geodesics exist under very generic conditions for
confining backgrounds.

\section{The Maldacena-N\'u\~nez background}

We begin by finding an appropriate null geodesic at $r=r_{0}$ in
the MN case.\footnote{A certain Penrose limit of the MN solution
was discussed in \cite{go}.}  This case is technically easier to
carry out, although it turns out to be more difficult to connect
with the dual field theory, due to a number of complicating
features.  It should be viewed, then, as a technical warm-up
exercise; we do not have a full understanding of its properties.

The MN background whose IR regime is associated with ${\cal N}=1$
SYM theory is that of a large number of D5 branes wrapping an
$S^2$. To be more precise: (i) the dual field theory to this
SUGRA background is the ${\cal N}=1$ SYM contaminated with KK
modes which cannot be de--coupled from the IR dynamics, (ii) the
IR regime is described by the SUGRA in the vicinity of the origin
where the $S^2$ shrinks to zero size.

The full MN SUGRA background includes the metric, the  dilaton and the
RR three-form.   In \cite{MN} an explicit expression of  the
background was written down based on   S-dualizing the background of
large N wrapped NS5 branes. The latter solution was constructed by
uplifting to ten dimensions an $SU(2)$ seven dimensional gauged SUGRA
for which the   spin connection of the $S^2$ is identified with the
$U(1)\in SU(2)$ gauge connection \cite{volkov}.

The background takes the following form
\bea
\label{IRmetric}
ds^2_{str} &= & e^{\phi_D } \left[ dx_\mu dx^\mu
+ \alpha' g_sN( d \rho^2 + e^{ 2 g(\rho)}
(d\theta_1^2+ \sin^2\theta_1 d\phi_1^2)+
{1 \over 4 } \sum_a (w^a - A^a)^2 ) \right]
\\
e^{2\phi_D} &=& e^{2\phi_{D,0}}{\sinh 2 \rho \over
 2 e^{g(\rho)} } \\
H^{RR}& =& g_sN \left[ - {1\over 4} (w^1 -A^1)\wedge (w^2 - A^2) \wedge ( w^3-A^3)  + { 1 \over 4}
\sum_a F^a \wedge (w^a -A^a) \right]
\eea
where $\mu=0,1,2,3$, we set the integration constant  $e^{\phi_{D_0}}=
\sqrt{g_s N}$. The expressions for $e^{2g(\rho)}$ and the gauge field $A$ are given by
\bea
\label{Afield}
e^{2 g}  &=& \rho \coth 2\rho - { \rho^2 \over \sinh^2 2 \rho } - { 1 \over 4} \\
A &=& { 1 \over 2} \left[ \sigma^1 a(\rho) d \t_1
+ \sigma^2 a(\rho) \sin\t_1 d\phi_1 +
\sigma^3 \cos\t_1 d \phi_1 \right]
\\
a(\rho) &= &{ 2 \rho \over \sinh 2 \rho}
\eea
  and the one-forms $w^a$ are given by:
\bea
 { i \over 2} w^a \sigma^a &  = &dg g^{-1}
\\
w^1 + i w^2 & = &e^{ - i \psi } ( d \t_2 + i \sin \t_2 d \phi_2)  ~,~~~~~~~~~~
w^3 = d \psi +\cos \t_2  d\phi_2  \\
g &=& e^{ i \psi \sigma^3 \over 2 } e^{ i \t_2 \sigma^1 \over 2 }
e^{ i \phi_2 \sigma^3 \over 2}
\\
\eea
Note that we use notation where $x^0,x^i$ have dimension of
length whereas $\rho$ and the angles
$\t_1,\phi_1,\t_2,\phi_2,\psi$ are dimensionless and hence the
appearance of the $\alpha'$ in front of the transverse part of
the metric. Moreover, following the notation of \cite{LS}
a factor of $g_s N$  is multiplying the $\alpha'$  instead of $N$ that
appears in \cite{MN}.

There are several scales associated with the ${\cal N}=1$  SYM dual of
the MN background. the string tension, the glueball masses, the
KK masses and the domain wall tension.  These masses are all
expressed in terms of the only scale of the background,
$\alpha'$, and they take the explicit form \cite{MN,LS}
\be
 M^2_{gb}\sim M^2_{KK}\sim {1\over g_s N\alpha'},\ \
 T_s \propto M^2_{gb}\,(g_s N)^{3\over 2}
\ee

\subsection{ The Penrose limit}
  We would like to take a Penrose limit for this
background following the general construction of section 2,
namely, based on  a null geodesic with $\rho=\rho_0$. In the
metric (\ref{IRmetric}) we can clearly see that $g_{tt}$ has a
minimum at $\rho=0$.  Here, the internal space in (\ref{IRmetric})
is an $S^3$; this suggest that motion at $\rho=0$ along an $S^3$
equator gives a candidate for a null geodesic. We would like to
solve for this null geodesic using a simplified metric of the
form (\ref{simple}).  In order to do this, we must switch to a
coordinate system where motion along the $S^3$ equator is
parameterized by a single angle, $\phi_+$, and such that the
description of the geodesic's neighborhood is particularly
simple.  Specifically we will ensure that any dependence of the
metric on the distance away from the chosen $S^3$ equator has to
be at least quadratic. This will guarantee that we can set the
first and second derivatives of any deviation to zero in the
equation of motion and solve for the null geodesic in terms of
just the variables $t,\rho$ and $\phi_+$.

 The coordinate system above is hard to find if we start from the explicit
form of the metric in (\ref{IRmetric}). There exists, however, a
simpler approach using the fact that at the origin the gauge
field (\ref{Afield}) is pure gauge, namely,
\be
iA=dh h^{-1}+O(\rho^2)\qquad \textrm{with}\qquad
h=e^{i\sigma^1\t_1/2}e^{i\sigma^3\phi_1/2}.
\ee
Performing a gauge transformation $A\rightarrow h^{-1}A h
+ih^{-1}dh $ sets the gauge field at the origin to zero up to
$O(\rho^2)$ corrections:
\begin{eqnarray}
\label{residual}
A &=& \Big(-{1\over 3}\rho^2 + {\mathcal{O}}(\rho^4)\Big)
\,\Big[\,\sigma^1(\cos\phi_1\,d\theta_1- \cos\t_1
\sin\t_1 \sin\phi_1\,d\phi_1 ) \nonumber\\
&&\; + \,\sigma^2(\sin\phi_1\,d\theta_1 + \cos\theta_1
\sin\theta_1 \cos\phi_1\,d\phi_1 ) +
\,\sigma^3(\sin^2\theta_1\,d\phi_1)\Big] .
\end{eqnarray}
Note that this is just
\be
\label{cylindrical}
A = - {1\over 3}\, [(r_2^2\, d\alpha_2)\,\sigma^1 + (r_1^2\,
d\alpha_1)\,\sigma^2 + (r_3^2\,d\alpha_3)\,\sigma^3]
\ee
where $(r_a,\alpha_a)$ are the
polar coordinates for the plane inside ${R}^3\sim { R}\times S^2$
which is perpendicular to the $x^a$ axis.
  It is straightforward to see that if we boost along a direction
$\omega^a$ on the $S^3$, the $A^a$ component of \eref{cylindrical}
will give the
only correction to the Penrose limit, proportional to
$r_a^2\,d\alpha_a\,dx^+$.  For example, if we choose to boost
along the great circle on $S^3$ defined by $\theta_2=0$ and
$\phi_2=\psi$ (hence boosting along $\omega^3$) and make the
following change of variables
\begin{eqnarray}
&&dt = dx^0,
 \qquad x^i \rightarrow {1\over L}\; x^i,
 \qquad \rho = {m_0\over L} \,r, \nonumber \\
&&\t_2 = {2\,m_0\over L}\,v ,\qquad \phi_+ = {1\over
2}(\ps+\phi_2),
\end{eqnarray}
where $L^2= \sqrt{g_s N}$ and
$m_0= {1\over \sqrt{g_s N \alpha'}}$ is the glueball mass, we get a limit for the metric
(\ref{IRmetric}) of the form:
\begin{eqnarray}
ds^2 & = & -\,L^2 dt^2 + \,dx_i dx^i
+\,dr^2 + r^2(d\t_1^2+\sin\t_1^2d \phi_1^2) \\
&&+ (dv^2+ v^2\,d\phi_2^2)
  + {L^2\over m_0^2}\,d\phi_+^2 - 2 v^2\,d\phi_2\,d\phi_+
  + {2\over 3}\,{r^2}\sin\theta_1^2\,d\phi_1d\phi_+ + {\mathcal{O}}(L^{-2})\nonumber
\end{eqnarray}
where the new variables $r,v$ have dimension of length.
It is  easy to  see that boosting along $\omega^1$
and $\omega^2$ for $\t_2 = {\pi\over 2}$ and $\psi = 0$ or
${\pi\over 2}$ will give the same type of result.
Since this form of the metric is invariant under $x\rightarrow
-x$, $x=0$ is a solution of the equation of motion. To eliminate
some of the ``magnetic" terms we now introduce a shift in the
angles $\phi_1$ and $\phi_2$
\be
\hat \phi_1 = \phi_1+{1\over 3}\,\phi_+\qquad \hat \phi_2=
\phi_2-\phi_+.
\ee
Expressed in terms of the these shifted angles the metric
takes the form
\begin{eqnarray}
ds^2 & = & L^2[-dt^2 +  {1\over m_0^2}\,d\phi_+^2] + dx_i dx^i \\
&&+\,{dr}^2
  + r^2(d\t_1^2+\sin\t_1^2d\hat\phi_1^2) \nonumber\\
&&+ (dv^2+ v^2\,d\hat\phi_2^2)
   -( {v^2}+
   {r^2\over 9}\sin^2\theta_1)\,d\phi_+^2 + {\mathcal{O}}(L^{-2}).\nonumber
\end{eqnarray}
Finally we let
\be
x^+=t, \qquad x^- = {L^2\over 2}(t-{1\over m_0}\,\phi_+),
\ee
and denote the Cartesian coordinates of the $R^3$ associated
with $dr^2 +r^2( d\t_1^2+\sin\t_1^2d \hat \phi_1^2)$ as $d u_1^2 +
du_2^2 + dz^2$ (and similarly for the $v$ plane
$dv^2 + v^2d\hat\phi_2^2=dv_1^2+dv_2^2$).  Then
 we take the Penrose limit
$ L \rightarrow \infty $  while keeping $m_0$ fixed, obtaining
\be\label{PLMN}
ds^2 = -2dx^+dx^- -m_0^2\,({1\over 9}u_1^2 + {1\over 9}
u_2^2+v^2)(dx^+)^2 + d\vec{x}^{\,2} +
d\vec{z}^{\,2} + du_1^2+du_2^2+dv_1^2+dv_2^2 \ .
\ee
We thus obtain a plane wave metric with 4 massless direction
(three $x$'s and $z$), two directions ($v$) with mass $m_0$
and two directions ($u$) with mass ${1\over 3}\,m_0$.

Next we would like to consider the  Penrose limit of the three  form
field strength  $H^{RR}$. According to  the  procedure of G\"uven
\cite{gueven}
since $H^{RR}$ is the field strength of the 2-form $A_2$,
  the $L\rightarrow\infty$ limit takes the form  $H^{RR} = L^2\tilde  H^{RR} $.
The only  terms that  survive this limit
are $ w^1 \wedge w^2 \wedge w^3$ and $F^3\wedge w^3$. All the other
terms are suppressed by factors of $1/L$. The final expression is
\be\label{HRR}
 H^{RR} = -2\,m_0\,dx^+\wedge [\,dv_1\wedge dv_2 + 1/3\, dz_1\wedge dz_2].
\ee
As expected the only non-trivial components are of the form $H_{+ij}$.
As a consistency check we examine the equation of motion
\be
R_{++}= {1\over 4}( H_{+ij}H_+^{ij}-{1\over 12}g_{++}H_{ijk}H^{ijk})
\ee
The component $R_{++}$ of the Ricci tensor associated with the
metric (\ref{PLMN}) is $R_{++}=\sum_i m_i^2 = (20/9)\, m_0^2$. It is
easy to see that by substituting  (\ref{HRR}) into the equation
of motion we get exactly the same expression also in the right
hand side of the equation. Notice that the second term in this
side of the equation vanishes since the terms of the 3-form have
a  structure of $H_{+ij}$ and $g^{++}=0$.

The Hamiltonian is:
\be
\label{MNH}
 H = -p_+ = i\pd_+ = E - m_0 (-{1\over 3} J_1 + J_2 + J_{\psi}) \equiv
E- m_0\,J,
\ee
and the momentum $P^+$ is
\be
P^+ = - {1\over 2} p_- = {i\over 2}\pd_- = {m_0 \over L^2}
(-{1\over 3} J_1 + J_2 + J_{\psi}) = m_0 {J\over \sqrt{g_s N}}.
\ee
where $J_1$, $J_2$ and $J_\psi$ denote $-i\pd_{\phi_1}$,
$-i\pd_{\phi_2}$ and  $-i\pd_{\psi}$ respectively.

Here we see something non-trivial, and slightly distressing,
about this plane--wave limit.  The metric \eref{IRmetric} with
the new gauge field \eref{residual} contains only two global
$U(1)$ isometries, $U(1)_L = J_2$ and $U(1)_R = J_\psi - J_1$
(modulo $SU(2)$ rotations). Therefore the symmetry current
$-{1\over 3} J_1 + J_2 + J_{\psi}$ does not represent an isometry
of the full MN solution; it is an accidental symmetry arising
only in the Penrose limit.  It therefore governs the hadrons of
the gauge theory only in the large-charge limit. Moreover, since
this accidental symmetry does not commute with supersymmetry, the
corresponding string theory will exhibit the supersymmetry of the
gauge theory in a slightly unexpected fashion.

  We could choose to define $\hat \phi_1 = \phi_1 + \phi_+$ so that
$J$ would now be in the diagonal of $U(1)_L$ and $U(1)_R$.
Unfortunately, this leaves us with a so-called "magnetic" term of
the form $d{\phi_1}d{\phi_+}$ in the metric.  Solving the string
theory in this background is slightly more complicated, but at
the end boils down to shifting the whole spectrum of energies
derived from the original background by ${2\over 3}m_0\ J_1$.

Perhaps one way to understand this strange appearance of a
magnetic term for what should be the natural choice of symmetry
current is to look at the dual field theory of the D5-brane. If we
look at the Kaluza-Klein spectrum for this D5-brane, we can
quickly see that it contains massive scalar and vector
multiplets.   The key feature here is that the scalar multiplets
transform under both of the global $SU(2)$'s, while the vectors
only transform under $SU(2)_R$! Capturing the full dynamics of the
Penrose limit requires us to look at objects with scalar
components from both the lowest mass scalar multiplet and the
lowest mass vector multiplet.   The $L\leftrightarrow R$
asymmetry of the vector scalars hints at a genesis for a magnetic
term.

Fortunately, the corresponding plane-wave limit of the KS solution
is not plagued with an un-natural choice of boost symmetry.
Although technically more challenging to obtain, it turns out to
be much more elegant and much easier to interpret.

\section{The Klebanov-Strassler background}

We begin by reviewing the KS background, which
is obtained by considering a collection of  $N$  regular
and $M$ fractional D3-branes in the geometry  of the deformed conifold
\cite{ks} (see also \cite{kh}).  The 10-d metric is of the form:

\be  ds^2_{10} =   h^{-1/2}(\tau)   dx_\mu dx^\mu  +  h^{1/2}(\tau) ds_6^2
\ ,  \ee  where $ds_6^2$ is the metric of the deformed conifold
\cite{candelas,mt}:

\be
\label{mtmetric}
ds_6^2 = {1\over 2}\varepsilon^{4/3} K(\tau)  \Bigg[ {1\over 3
K^3(\tau)} (d\tau^2 + (g^5)^2)  +  \cosh^2 \left({\tau\over
2}\right) [(g^3)^2 + (g^4)^2]  + \sinh^2 \left({\tau\over
2}\right)  [(g^1)^2 + (g^2)^2] \Bigg].
\ee
where
\be
K(\tau)= { (\sinh (2\tau) -
2\tau)^{1/3}\over 2^{1/3} \sinh \tau},
\ee
and
\bea
\label{forms}
g^1 &=&
{1\over \sqrt{2}}\big[- \sin\theta_1 d\phi_1  -\cos\psi\sin\theta_2
d\phi_2 + \sin\psi d\theta_2\big] ,\nonumber \\  g^2 &=& {1\over
\sqrt{2}}\big[ d\theta_1-  \sin\psi\sin\theta_2 d\phi_2-\cos\psi
d\theta_2\big] , \nonumber \\  g^3 &=& {1\over \sqrt{2}} \big[-
\sin\theta_1 d\phi_1+  \cos\psi\sin\theta_2 d\phi_2-\sin\psi d\theta_2
\big],\nonumber \\  g^4 &=& {1\over \sqrt{2}} \big[ d\theta_1\ +
\sin\psi\sin\theta_2 d\phi_2+\cos\psi d\theta_2\ \big],   \nonumber
\\  g^5 &=& d\psi + \cos\theta_1 d\phi_1+ \cos\theta_2 d\phi_2.
\eea
The 3-form fields are:
\begin{eqnarray}
F_3 &=& {M\alpha'\over 2} \left \{g^5\wedge g^3\wedge g^4 + d [
F(\tau)  (g^1\wedge g^3 + g^2\wedge g^4)]\right \} \nonumber \\  &=&
{M\alpha'\over 2} \left \{g^5\wedge g^3\wedge g^4 (1- F)  + g^5\wedge
g^1\wedge g^2 F \right. \nonumber \\  && \qquad \qquad \left. + F'
d\tau\wedge  (g^1\wedge g^3 + g^2\wedge g^4) \right \}\ ,
\end{eqnarray}
and  \be  B_2 = {g_s M \alpha'\over 2} [f(\tau) g^1\wedge g^2  +
k(\tau) g^3\wedge g^4 ]\ ,  \ee
\begin{eqnarray}
H_3 = dB_2 &=& {g_s M \alpha'\over 2} \bigg[  d\tau\wedge (f'
g^1\wedge g^2  +  k' g^3\wedge g^4)  \nonumber \\  && \left. + {1\over
2} (k-f)  g^5\wedge (g^1\wedge g^3 + g^2\wedge g^4) \right]\ .
\end{eqnarray}
The self-dual 5-form field strength is  decomposed as $\tilde F_5 =
{\cal F}_5 + \star {\cal F}_5$, with  \be  {\cal F}_5 = B_2\wedge F_3
= {g_s M^2 (\alpha')^2\over 4} \ell(\tau)  g^1\wedge g^2\wedge
g^3\wedge g^4\wedge g^5\ ,  \ee  where  \be  \ell = f(1-F) + k F\ ,
\ee  and  \be  \star {\cal F}_5 = 4 g_s M^2 (\alpha')^2
\varepsilon^{-8/3}  dx^0\wedge dx^1\wedge dx^2\wedge dx^3  \wedge
d\tau {\ell(\tau)\over K^2 h^2 \sinh^2 (\tau)}\ .  \ee  The functions
introduced in defining the form fields are:  \bea  F(\tau) &=& {\sinh
\tau -\tau\over 2\sinh\tau}\ ,  \nonumber \\  f(\tau) &=&
{\tau\coth\tau - 1\over 2\sinh\tau}(\cosh\tau-1) \ ,  \nonumber \\
k(\tau) &=& {\tau\coth\tau - 1\over 2\sinh\tau}(\cosh\tau+1)  \ .
\eea  The equation for the warp factor is  \be \label{firstgrav}  h' =
- \alpha {f(1-F) + kF\over K^2 (\tau) \sinh^2 \tau}  \ ,  \ee  where
\be  \alpha =4 (g_s M \alpha')^2  \varepsilon^{-8/3}\ .  \ee
For large $\tau$ we impose the boundary condition that  $h$
vanishes. The resulting integral expression for $h$ is  \be
\label{intsol}  h(\tau) = \alpha { 2^{2/3}\over 4} I(\tau) =  (g_s
M\alpha')^2 2^{2/3} \varepsilon^{-8/3} I(\tau)\ ,  \ee  where  \be
I(\tau) \equiv  \int_\tau^\infty d x {x\coth x-1\over \sinh^2 x}
(\sinh (2x) - 2x)^{1/3}  \ .  \ee  The above integral has the
following expansion in the IR:  \be  I(\tau\to 0) \to a_0 - a_1
\tau^2 + {\cal O}(\tau^4) \ ,  \ee  where $a_0\approx 0.71805$ and
$a_1=2^{2/3}\, 3^{2/3}/18$. The absence  of a linear term in $\tau$
reassures us that we are really expanding  around the end of space,
where the Wilson loop will find it more favorable to arrange itself.

\subsection{A convenient parameterization of the KS background}

As it turns out, the above parameterization of the metric will not
be quite suitable for our purposes since the 1--forms $d\psi$,
$d\theta_i$ and $d\phi_i$ mix the 1--forms of the $S^3$ at the
origin, $g^3,g^4$ and $g^5$, with those from the $S^2$, $g^1$ and
$g^2$.  In these coordinates it would be problematic to try and
get a Penrose limit by boosting along $g^5$. For example,
boosting along $g^5$ by shifting $\psi$ does not work as this
coordinate is Hopf--fibered over the $S^2$ which shrinks to zero
size at the origin.

    Instead, we pick explicitly separate coordinates\foot{This is
very similar in spirit to the gauge transformation we used in the
last section.} for the $S^3$ and the $S^2$ (see the appendix for
more detail) : for the $S^3$ an $SU(2)$ matrix
\be
T  = e^{{i\over 2}\,\phi'\, \sigma_3}\,e^{{i\over 2}\,\theta'\,
 \sigma_1} \,e^{{i\over 2}\,\psi'\,
 \sigma_3},
\ee
and for the $S^2$ a matrix
\be
 S = e^{{i\over 2}\,\phi\, \sigma_3}\,e^{-{i\over 2}\,\theta\,
 \sigma_1}.
\ee
We can now work with the 1--forms
\be
T^\dag\,dT = - dT^\dag\,T ={i\over 2}\,\omega^a\,\sigma_a,
\ee
and
\be
 d\theta\;,\;  \sin\theta\, d\phi.
\ee
We now re-write the metric for the deformed conifold in terms of
these 1--forms as
\bea
\label{newmetric}
\ep^{-{4\over 3}}\,ds_6^2 &=&
 {1\over 4}\,K(\tau)\,\cosh(\tau)\,\left(d\tau^2 + (\omega^a)^2\right) \nonumber \\
 &+& \,K(\tau)\,\sinh^2({\tau\over 2})\;
 \Bigg[\;
 (d\theta^2 + \sin^2\theta\,d\phi^2) \\
 &&\qquad - \,(\sin\phi\,\omega^1 + \cos\phi\,\omega^2)(d\theta) \nonumber\\
 &&\qquad - \,(\cos\theta\cos\phi\,\omega^1 - \cos\theta\sin\phi\,\omega^2
 - \sin\theta\,\omega^3)(\sin\theta\,d\phi) \nonumber
 \Bigg] \nonumber \\
&+&
 {1\over 4}\,K'(\tau)\,\sinh(\tau) \left[\;d\tau^2\; +
 \; (\sin\theta\cos\phi\,\omega^1 +
 \sin\theta\sin\phi\,\omega^2
 +\cos\theta\,\omega^3)^2\right].  \nonumber
\eea

\subsection{Plane wave limit}
Due to the behavior of the warp factor in the IR
$(h\to\rm{constant})$, it is clear that in the deep IR there are
null geodesics that lie at $\tau = 0$. Hence, we will consider a
Penrose limit where we expand around $\tau=0$, in a manner
similar to the BMN expansion near the center of $AdS$ in global
coordinates. (The analogy here is purely formal, however, as the
physical meaning of the time variable in global versus
Poincar\'{e} coordinates is very different in the field theory
dual.)  An important guide in taking the limit that we want is that
we should keep finite the mass of the glueball
\be
M_{gb} \propto {\varepsilon^{2\over 3}\over g_s\,M\,\alpha'}.
\ee
Note that now the dynamics in KS are such that the flux tube tension is
\be
T_s \propto M_{gb}^2 (g_s M).
\ee
We start the machinery for the Penrose limit by expanding the KS
metric up to quadratic terms in $\tau/L$, and eventually taking
$L\to \infty$.  We also want to take a Penrose limit near an
equator on the $S^3$ at the origin.  Without loss of generality,
we can choose coordinates $\theta',\phi'$ and $\psi'$ such that
this equator sits at $\theta'=0$ and is generated by
$\phi'+\psi'$ (to first order this is $\omega^3$).  To take the
limit, we will need also need to re-scale the coordinate $\theta'
\to {\theta'/L}$, i.e., the 1--forms $\omega^1, \omega^2$ will go
like ${1\over L}$. This scaling simplifies the deformed conifold
metric (\ref{newmetric}):
\bea
  L^2 ds_6^2 &=& { \ep^{4\over 3} \over  2^{5\over 3} \;3^{1\over 3}}
             \Bigg[\,d\tau^2
         + (d\theta')^2 + (\theta')^2(d\phi'^2) + L^2 (\omega^3)^2
         + {2\over 5}\tau^2\, (\omega^3)^2  \nonumber\\
         &+& \tau^2 d\Omega_2^2 +
         (\tau^2\,\sin^2\theta)d\phi\,\omega^3 - {1\over 5}
         \tau^2 \cos^2\theta (\omega^3)^2 \Bigg].
\eea
If we expand $\omega_3^2$ as
\be
 \omega_3^2 = 4\,(d\phi_+)^2 -
 2\,\left({\theta'\over L}\right)^2d\phi'\,d\phi_+,
\ee
we can write the full 10--dimensional metric in the limit as:
\bea
 ds_{10}^2 &=& -{c_0^2\over L^2} \bigg[L^2+{a_1\over 2a_0}\tau^2 \bigg] dt^2
  + c_0^2 \delta_{ij}dx^i dx^j  \\
 &+& {c_1^2\over L^2}\bigg[4 L^2 d\phi_+^2 + d\tau^2 +
 \tau^2\,(d\theta^2 + \sin^2\theta\,d\phi^2)
  + (d\theta')^2 + (\theta')^2(d\phi')^2
   \nonumber \\
 &+& 2 \tau^2\,\sin^2\theta\,d\phi\,d\phi_+
  -  2 \theta^2\,d\phi'\,d\phi_+ \nonumber\\
 &+& 4 \tau^2\sin^2\theta\left({2\over 5}-{a_1\over 2a_0}\right)(d\phi_+)^2
  + 4 \,\tau^2\cos^2\theta\left({1\over 5}-{a_1\over 2a_0}\right)(d\phi_+)^2\bigg],
  \nonumber
\eea
with
\be  c_0^2
={\varepsilon^{4/3}\over 2^{1/3}\,\,g_s \,\,M \alpha'
\,\,a_0^{1/2}}, \qquad c_1^2 = {g_s\,\,M\,\,\alpha'
\,\,a_0^{1/2}\over 2^{4/3}\,\,3^{1/3}} .
\ee
As a next step, we know that the Penrose limit calls for the
overall metric to be re-scaled by $L^2$.  We can accomplish this
by taking
\be
c_0 \rightarrow \infty,\qquad c_1 \rightarrow\infty,
\ee
while keeping constant
\be
{c_0\over L} = 1,
 \qquad {c_0 \over c_1} = {\varepsilon^{2\over 3}\over g_s M \alpha'}
 \Bigg({24\over a_0^3}\Bigg)^{1\over 6}= 2{m_0}.
\ee
With these scalings in mind, we make the following further changes
in coordinates in order to take the Penrose limit:
\be
 \phi_+ = {1\over 2}(\phi'+\psi'),
 \qquad x^+ = t,
 \qquad x^{-}= {c_0^2\over 2} \left(t - {2 c_1\over c_0} \,{\phi_+ }\right),
\ee
with
\bea
 x^i \to {x^i \over L},&& \quad \varphi = {1 \over 2 }(\phi' - \psi'),
 \quad v = {c_1\over c_0}\,\theta'\,e^{i\varphi}, \\
 z = {c_1\over c_0} \,\tau\cos\theta\, && \quad \tilde{\phi} = \phi
 + \,\phi_+,\quad u = {c_1\over c_0}\,\tau\sin\theta\,e^{i\tilde{\phi}}
 . \nonumber
\eea
After we take $L \to \infty$ the resulting metric is
\bea
\label{metks}
 ds^2 = -4dx^+dx^-
 &-& m_0^2\bigg[\,\left({4a_1\over a_0}-{4\over 5}\right)\, z^2 +
 \left({4a_1\over a_0}-{3\over 5}\right)\,u\bar{u} + v\bar{v}\bigg] \, (dx^+)^2 \nonumber \\
 &+& dx^i dx_i + dz^2 + dud\bar{u} + dvd\bar{v}.
\eea

\subsection{The various forms in the plane wave limit}

We now turn to the construction of the forms in the new coordinates.
A convenient relation we will use in what follows is:
\be
  {g_sM\alpha'\over L^2}\,m_0^2 = a_0^{-{1\over 2}} 2^{{4\over 3}}
  3^{{1\over 3}} {c_1^2\over c_0^2} \, m_0^2 = a_0^{-{1\over 2}}2^{-{2\over 3}}
  3^{{1\over 3}} = \left({a_1\over a_0}\right)^{1\over 2}
  {3\over\sqrt{2}}.
\ee

The earlier expressions for the $g^i$'s (\ref{forms}) allow us to write
down the Ramond--Ramond 3--form:
\begin{eqnarray}
F_3 &=& {M\alpha'\over 2} \left\{g^5\wedge g^3\wedge g^4 + d \left[
      F(\tau)  \left(g^1\wedge g^3 + g^2\wedge g^4\right)\right]\right\} \\
    &\to& {3i\,m_0\over \sqrt{2}\,g_s} \left({a_1\over
    a_0}\right)^{1\over 2} dx^+\wedge\left(\,{1\over
    3}\, du\wedge d\bar{u} + dv\wedge d\bar{v}\right). \nonumber
\end{eqnarray}
Similarly we write down the NS--NS 2--form as:
\bea
\label{bks}
B_2 &=& {g_s M \alpha'\over 2}\, [f(\tau) g^1\wedge g^2 + k(\tau)
          g^3\wedge g^4 ] \\
    &\to& {m_0\over \sqrt{2}}\left({a_1\over
    a_0}\right)^{1\over 2}\,dx^+\wedge
    (-i)[u\, d\bar{v} - \bar{u}\, dv] \nonumber
\eea
The complex 3-form field strength obtained by combining the above forms is:
\bea
\label{G3}
G_3 &=& H_3 + ig_s F_3 \\
    &=& {m_0\over \sqrt{2}}\left({a_1\over
    a_0}\right)^{1\over 2}\,dx^+\wedge\left[
    (\,du\wedge d\bar{u}+ 3 \, dv\wedge d\bar{v})
    + i (du\wedge d\bar{v} - d\bar{u}\wedge dv)\right],
    \nonumber
\eea
which has as a norm
\be
(G_3)_{+ij}\,(\overline{G}_3)_{+}^{\;\,ij}
 = 48 \,{a_1\over a_0}\, m_0^2.
\ee
As an extra check we verify that the only nontrivial equation of motion
\be
R_{++} = {1\over 4}
  (G_3)_{+ij}\,(\overline{G}_3)_{+}^{\;\,ij}
\ee
is satisfied. Indeed from (\ref{metks}) we obtain
\be
R_{++} = m_0^2\left[\,\left({4a_1\over a_0}-{4\over 5}) +
 2 ({4a_1\over a_0}-{3\over 5}\right) + 2\right] =  12\,{a_1\over a_0}\,
 m_0^2,
\ee
which matches perfectly with the 3--form.

\subsection{Operators and symmetries}

The Hamiltonian now takes the
form
\bea
\label{KSH}
 H=-\,p_+&=&i\pd_{+}=i\big[\pd_t  + m_0 \,(\pd_{\phi'}+
\pd_{\psi'} - \pd_{\phi}) \big] \nonumber\\
&=&E-m_0J,
\eea
with
\bea P^+&=&{i\over 2} \pd_{-}= - {i\over c_0^2}\, m_0\,(\pd_{\phi'}+
\pd_{\psi'} - \pd_{\phi})\nonumber \\
 &=& m_0\,\left(\,{J \over
c_0^2}\,\right).
\eea

%
%
The geodesic used in our Penrose limit is generated by
a symmetry which we will call $U(1)_D$.  Its action
on the matrices $T$ and $S$ is
\bea
e^{i\alpha\,J}:  T &\to& e^{i{\alpha\over 2}\,\sigma_3}\,T\,
e^{i{\alpha\over 2}\,\sigma_3} \nonumber \\
                 S &\to& e^{-i{\alpha\over 2}\,\sigma_3}\,S
\eea
which means its action on our general complex
coordinate matrix for the conifold is
 $W = T\,S\,W_{\ep}\,\sigma_3\,S^{\dag}\sigma_3,$ is:
\be
e^{i\alpha\,J}: W \to e^{i{\alpha\over 2}\,\sigma_3} W
e^{i{\alpha\over 2}\,\sigma_3}.
\ee
The geodesic
is left invariant by an orthogonal abelian
symmetry acting on $W$, which we will term $U(1)_A$,
with
\be
 e^{i\alpha\,J_A}: W \to e^{i{\alpha\over 2}\,\sigma_3} W
 e^{-i{\alpha\over 2}\,\sigma_3}.
\ee
In this case,
\be
J_A = -i(\pd_{\phi'} - \pd_{\psi'} + \pd_{\phi}) \ .
\ee
One may check that both $u$ and $v$ carry charge $1$ under
$J_A$, while $x^+$ and $z$ are neutral, and that the metric and
3-forms are also neutral.

Thus, in contrast to our results in the MN case, the symmetry current
corresponding to the charge $J$ is conserved by the full gauge theory,
and corresponds to an isometry of the full KS metric. Moreover,
it commutes with the supersymmetry of the full theory.  This makes
the corresponding string theory, and its interpretation, reasonably
straightforward.

\section{The plane wave  string  and its Hamiltonian}

In this section we work out the string Hamiltonian for the KS and MN
plane waves, keeping an eye out for common distinguishing features as
well as differences.  We will treat the bosonic sector first, and then
discuss the fermionic oscillators and the effects of supersymmetry.

\subsection{Bosonic Sector}

 The form of the KS metric (\ref{metks}) directly implies that the
bosonic sector of the system is described by three massless fields
with frequencies $w_n=n$, and five massive (no zero--frequency
mode) fields.  Due to the presence of a B-field, four of the
latter organize themselves as two sets of coupled fields (see
appendix B).  The frequencies for the five massive fields are
\bea
\label{freq}
w_n^z &=& \sqrt{n^2 + \hat m_z^2} \\
(\omega_n^{\pm})^2 &=&{1\over 2}\bigg[2\,n^2+\hat m_v^2 +\hat m_u^2 \pm
\sqrt{(\hat m^2_v-\hat m_u^2)^2+4\,n^2\,\hat m_B^2} \bigg],\nonumber
\eea
where
\bea
\hat m_z = (m_0 p^+\a')\left({4a_1\over a_0}-{4\over 5}\right)^{1\over 2},
\quad &&
\hat m_u = (m_0 p^+\a')\left({4a_1\over a_0}-{3\over 5}\right)^{1\over 2},
\quad \\
\hat m_v=m_0 p^+\a', \quad &&
\hat m_B=\sqrt{2}m_0 p^+ \a'({a_1\over a_0})^{1/2}. \nonumber
\eea
Glancing at the form of the frequencies for the coupled fields we
notice that the presence of the B-field (parameterized by $\hat
m_B$ above) only affects the frequencies for $n>0$. At zero-level
the coupled fields have frequencies $\omega_0^+ = \hat m_v$ and
$\omega_0^- = \hat m_u$ which can naturally be identified with
excitations of $v$ and $u$.

 We can obtain the structure for the lowest-lying excitations of the
MN plane wave case in a similar fashion to the case above.  Due to the
absence of a B--field in the MN plane--wave the directions $u$ and $v$
no-longer mix, so we can recycle formula in eq. \eref{freq} with
\be
\hat m_z = 0,
\quad
\hat m_u = {1\over 3}(m_0 p^+\a'),
\quad
\hat m_v=m_0 p^+\a', \quad
\hat m_B=0.
\ee
and rename
\be
\omega_n^+ = \omega^v_n,\qquad \omega_n^- = \omega^u_n.
\ee

What do the MN and KS spectrum have in common?  First, they have
an identical stringy sector for the three directions which
correspond to the three spatial directions in the dual gauge
theory.  Second, they have the same massive level-zero modes in
two of the internal directions ($v$~and~$\bar v$).  These come
from a natural $S^3$ structure in both the MN and KS cases; they
represent the direction in this sphere normal to the reference
geodesic.  Even though this common feature is ruined by the
B--field for more excited states, we will refer to oscillation in
$v$ as the ``universal sector.''  The remaining oscillation
directions are less universal and come from the combination of
the radial direction and the $S^2$-like structure.  In
particular, what is striking here is that $\hat m_z$ is zero in
the MN case, while all the $\hat m$'s are positive in the KS
case.  In either case, it is interesting to note that all the
quantities in this ``non-universal sector'' are smaller than
$\hat m_v$ since $\sqrt{{4a_1\over a_0}- {4\over 5}}\approx .47$
and $\sqrt{{4a_1\over a_0}-{3\over 5}}\,\approx .65$.

The whole bosonic Hamiltonian can be written explicitly following standard manipulations. Here we provide the needed notation to understand its form; the details are given in appendix B.
We define number operators
\bea
\N_R&=&\sum_{n=1}^\infty n \big( a^{i\dagger}_{n} a_{n}^i \big) \
,\ \ \ \ \N_L=\sum_{n=1}^\infty n \big( \tilde a^{i\dagger}_{n}
\tilde a^i_{n}
\big)\nonumber \\
N_R &=&\sum_{n=1}^\infty n \big( a^{s\dagger}_{n} a^s_{n} \big) \
, \ \ \ \ N_L=\sum_{n=1}^\infty n \big( \tilde a^{s\dagger}_{n}
\tilde a^s_{n} \big)\ ,
\eea
and sub--Hamiltonians
\bea
H_0&=&  w_0^s \big(a^{s\dagger}_{0} a_{0}^s
\big) \ , \nonumber \\
H_R &=& \sum_{n=1}^\infty w_n^s \big(a^{s\dagger}_{n} a_{n}^s \big),\,\,\,
H_L=\sum_{n=1}^\infty w_n^s \big( \tilde{a}^{s\dagger}_{n}
\tilde{a}_{n}^s \big) \ .
\eea
The subindex $i=1,2,3$ refers to the three flat directions in the
plane wave (spatial directions in the gauge theory), while the
index $s=4,5,6,7,8$ runs over the internal directions. There is
implied summation over the indices $i$ and $s$. The full bosonic
light-cone Hamiltonian is
\bea
\label{TheHamiltonian}
H=-P^- &=& H_{\parallel} + H_{\perp} \\
&=& \left[{P_{i}^2 \over 2P^+ } + {1\over 2 \a' P^+}\left(\N_R+
\N_L\right)\right]
+ \left[{1\over 2 \a' P^+}\left(H_0+ H_R+ H_L  \right)\right]
\nonumber\ .
\eea
 The Hamiltonian is thus constructed of
a contribution from the momentum and stringy excitations in the
spatial directions of the field theory (index $i$),
$H_{\parallel}$, and a contribution from the massive ``zero''
modes and excitations of the internal directions (index $s$),
$H_{\perp}$.

From this we may observe two important features which both the
bosonic MN and KS Hamiltonians share. First, both theories have
the same $H_{\parallel}$. Second, they have the same $\hat m_v =
p^+\alpha' m_0$.  More precisely, note from \eref{MNH} and
\eref{KSH} that we have defined $m_0$ in each case so that the
energy $E$ of the string theory vacuum state is $Jm_0$. The two
theories then share the fact that the lowest-lying mode of $v$
shifts $E$ by exactly $m_0$.  We will see in a moment why these
features are universal.

\subsection{The fermionic sector}

   Before we describe the fermionic contribution to the string spectrum in our
plane waves, let us consider first what happens to the target
space supersymmetries of the original solutions.  Both the MN and
KS background are dual to $\N=1$ supersymmetric field theories;
their supersymmetry algebra contains exactly four supercharges.
These supercharges commute with the original Hamltonian, $\pd_t$,
and with the $SU(2)_L\times SU(2)_R$ global symmetry generators.
In the Penrose limit, these supercharges are re--scaled as
\be
Q \to {L}\cdot Q
\ee
since their Killing spinors mix with the coordinates $x^i$.
This implies a contraction of the supersymmetry algebra
\bea
\{Q,Q\} &\propto& \Gamma^\mu {\mathcal P}_\mu \\
\to \qquad \{Q,Q\} &\propto& {1\over L^2} \left(\Gamma^0
(i\pd_+\, + \, L^2\, i\pd_-) + L\,\Gamma^i i\pd_i \right) =
\Gamma^0 P^+ + {\mathcal O}(L^{-1}),\nonumber
\eea
which tells us that the original supercharges have now become
part of the 16 ``kinematic" supercharges ubiquitous to pp--wave
solutions (\cite{radu,cvetic1}).  This means (see
\cite{cveticetal}) that they will be non-linearly realized on the
string worldsheet.

Let us now specialize to the KS case.  The Hamiltonian for the KS
plane--wave is shifted from the original KS Hamiltonian by a
charge J which generates $U(1)_D$ in the global symmetry group.
This means that it still commutes with the four supercharges
above!  After fixing lightcone gauge and kappa symmetry, the
(non-linear) action of the kinematic supersymmetries takes the
form of multiplication by the Green-Schwarz fermionic fields
$S^\alpha$ and $\tilde S^{\alpha}$.  Commutation of four of the
sixteen supersymmetries with the Hamiltonian then implies that
two each of the eight left--moving and right--moving spinors on
the worldsheet should have a zero--frequency mode. These act on
the vacuum to generate a four-dimensional Hilbert space of degenerate
states (two fermionic and two bosonic).

An explicit computation (see Appendix B) of the fermionic
spectrum confirms this prediction. The spectrum of the string in
the KS plane wave contains four fermionic fields (each of which
has a left--moving and a right--moving part) with frequencies
\be
\omega_n^{\alpha}|_{\alpha = 1...4} = \sqrt{n^2 + \hat m_f^2},
\ee
two fermionic fields with frequencies
\be
\omega_n^{\alpha}|_{\alpha = 5,6} = \sqrt{n^2 + {1\over 2} \hat m_f^2 +
{1\over 2}{\hat m_f}\sqrt{\hat m_f^2 + 4n^2}},
\ee
and two fermionic fields with frequencies
\be
\omega_n^{\alpha}|_{\alpha = 7,8} = \sqrt{n^2 + {1\over 2} \hat
m_f^2 - {1\over 2}{\hat m_f}\sqrt{\hat m_f^2 + 4n^2}}.
\ee
The mass scale is $\hat m_f = m_0\, \left({2 a_1\over
a_0}\right)^{1\over 2}\,p^+ \alpha'$. For $n=0$ we have six modes
with frequency $\hat m_f$ and then we get the two zero--modes we
expected.

  Let us now look at the MN case.  For this background, the spectrum
of fermionic oscillators is much easier to compute.  These
oscillators come in two sets of four with frequencies
\bea
\label{MNferm}
\omega_n^{++\beta} = \omega_n^{--\beta} &=& \sqrt{n^2 + {4\over 9} (m_0p^+\alpha ')^2}, \\
\omega_n^{-+\beta} = \omega_n^{+-\beta} &=& \sqrt{n^2 + {1\over 9} (m_0p^+\alpha')^2}.\nonumber
\eea
where $\beta = 1$ or $2$ and the signs $(\pm\pm)$ represent
eigenvalues $\pm{1\over2}$ under rotations in the $u$ and $v$
planes respectively.  We have chosen to label the fermionic
oscillations via these eigenvalues to illustrate a subtle yet
simple point about the action of the original susy's from the MN
solutions.

  As we noted at the end of Sec.~3, the current $J = -{1\over 3}J_1 +
J_2 + J_{\psi}$ used in the MN plane-wave limit is not in the global
symmetry group $SU(2)\times SU(2)$ of the full theory.  Correspondingly the
Hamiltonian for the corresponding plane wave does not commute with the
original four $N=1$ supercharges.  On the other hand, the operator $J'
= J - {2\over 3} J_1 = J - {2\over 3}J_u$ {\em is} an element of the
original global symmetry group, so if we shift the Hamiltonian by
$-{2\over 3}J_u$ we should get two zero frequency modes.  Taking a
careful look at \eref{MNferm} we see that this shift takes
\be
\omega_n^{+-\beta} \to \sqrt{n^2 + {1\over 9} (m_0p^+\alpha')^2}-
{1\over 3}(m_0p^+\alpha')
 \ee
giving the two required zero--modes for $n=0$.  Thus, our string theory
does exhibit the supersymmetry of the field theory, but it makes it
somewhat hard to see.

Now that we described the spectrum of the fermionic Hamiltonian
for MN and KS, we should make clear a few important connections
with the bosonic Hamiltonian.  First, if we define fermionic
number operators
\be
 N_L^f = \sum_{n=1}^\infty n \big( S^{\alpha\dagger}_{n}
S^{\alpha}_{n}) \ ,\qquad
 N_R^f = \sum_{n=1}^\infty n \big(
\tilde S^{\alpha\dagger}_{n} \tilde S^{\alpha}_{n}) \ ,
\ee
the contribution of the bosonic and fermionic modes is constrained
by the equality of the overall occupation numbers \cite{russo}
which have to satisfy the level-matching condition $N_R+ \N_R +
N^f_R = N_L+ \N_L + N^f_L$.  Second, we note that in both cases
the sum of the squares of the fermionic frequencies above exactly
matches the sum of the squares of the frequencies of the bosonic
fields in \eref{freq} order by order in $n$.  This allows the
corresponding string-theory to remain finite.  Finally, neither
the MN case nor the KS case has any linearly--realized worldsheet
supersymmetries.  This implies that there will be a zero--point
energy for the overall Hamiltonian.  Since at each level the sum
of the fermionic frequencies is bigger than the sum of the
bosonic frequencies, we will get a positive zero-point energy;
there is no tachyon.

  To conclude, the fermionic contribution to the Hamiltonian
incorporates quite well our knowledge of the supersymmetries,
especially for KS.  The MN and KS pp-wave string theories are
solvable, finite, and built on a positive--energy vacuum.

\section{A string theory of hadrons}

 In order to interpret the Hamiltonian above in terms of the field theory dual
to the ``parent" background we must keep in mind the following
facts.   Local inertial momenta $P_{i}$, as measured in the string frame
near $r = r_0$, are related to momenta in the
 field theory, ${\mathcal P}_{i}$, via the relation
\be
{\mathcal P}_{i} = g_{ii}(r_0) P_{i}.
\ee
We can also write the confined theory string tension, $T_s$, in
terms of the string length as
\be
T_s = (g_{tt}(r_0)\, g_{xx}(r_0))^{1\over 2} {1\over \a'} = g_{tt}(r_0)
{1\over \a'}.
\ee
Now eq. (\ref{TheHamiltonian}) can be written purely in terms of
field theory variables as:
\be
\label{TheFieldTheoryHamiltonian}
H = \left[{{\mathcal P}_{i}^2 \over 2m_0 J } + {T_s\over 2 m_0 J}
\left(\N_R+ \N_L\right)\right] + \left[{T_s\over 2 m_0 J}
\left(H_0+ H_R+ H_L  \right) \right]\ .
\ee

\subsection{The toy model of a string on a compact circle}

Before discussing the interpretation of
these hadrons in the KS and MN theories, we begin by noting
that there is a simple toy model\footnote{M.J.S.
thanks Minxin Huang, Thomas Levi, and Asad Naqvi for
discussions concerning this toy model prior to the present
work.} ---  a string moving on a compact
circle --- which shares some parts of this Hamiltonian.
As such, it helps to orient us toward a clear interpretation
of the physics, although it does not capture all of the features
of the Hamiltonian in \eref{TheFieldTheoryHamiltonian}.
We simply consider a closed unwound
string on flat ${\cal M}^9\times S^1$,
the circle having radius $R_0$.

First
consider an excited string at rest.  Its energy is
\be
\sqrt{{1\over \alpha'} ( N_L + N_R)}
\ee
where (ignoring worldsheet fermions)
\be
 N_R =\sum_{n=1}^\infty n \big( a^{i\dagger}_{n} a^i_{n}
 \big) \ , \ \ \ \ N_L=\sum_{n=1}^\infty n
\big( \tilde a^{i\dagger}_{n} \tilde a^i_{n} \big)\ , \nonumber \\
\ee
except that we sum over all 8 noncompact directions $x^1,\dots, x^8$
transverse to a light cone (which
 we place in the  directions $x^0,x^9$).

Now let us boost the string, giving it small momentum $\vec {\bf P}$ in three
spatial Minkowski
directions and enormous
momentum $P_9\equiv J/R$ in the $x^9$ direction.  Its energy is now
\be
\sqrt{P_9^2 + \vec {\bf P}^2 + {1\over \alpha'} (N_L + N_R)}
\ee
and so
\be
E-P_9 = {\vec {\bf P}^2\over 2J/R_0} +
{1\over 2J/R_0}
 {1\over \alpha'} \left(N_L + N_R\right)
\ee
which looks similar to the formula
\eref{TheFieldTheoryHamiltonian} above if we identify $R_0$ as
$1/m_0$, and $T_s$ as ${1\over\a'}$.

How should we interpret this?  From the ten-dimensional point of view,
this is merely Lorentzian physics.  But from the {\it
nine-dimensional} point of view, we are adding not momentum but {\it
KK charge} to the string, whereby it remains static but becomes heavy.
Any additional motion of the string in the noncompact directions looks
like nonrelativistic motion from the point of view of nine dimensions.
Excitations of the boosted string, which look perfectly ordinary from
the ten-dimensional point of view, take the above squared form from
the nine-dimensional point of view.  From this we learn that the first
terms in \eref{TheHamiltonian} and \eref{TheFieldTheoryHamiltonian}
simply reflect how large and heavy
nonrelativistic strings move and oscillate.\footnote{Other related toy models
can easily be found; for example, one might consider lifting the toy
model to M-theory, and through an 11-9 flip relating the oscillations
on a
boosted string to strings on a bound state of $D_0$ branes.  Indeed
the Hamiltonian for excitations of such a bound state will look very
similar to the first terms in our Hamiltonian. In all cases, it
is the effect of tacking on a small oscillation to a large mass by
addition in quadrature.  Indeed Hamiltonians of this type
have appeared many times
in the contexts of DLCQ and strings with large winding number.}

Thus we are led to guess that the hadrons described in our string
theory take the physical shape of nonrelativistic strings
propagating in four dimensions.\footnote{More precisely, highly
excited hadrons in our string theory actually ``look'' like
strings. Low-lying states are small, essentially point particles,
in the same way that gravitons in ordinary string theory do not
look like strings but instead have well-localized wave
functions.  The wave functions for our low-lying hadrons can be
guessed from those of ordinary strings, using the toy model.}
Since these hadrons have never been studied before, they need a
name: we will call them ``{\it annulons}'' from the Latin word
``anulus'' for ``ring.''  The vacuum of the string theory is the
lowest-lying, stable annulon with charge $J$, and our string
theory describes its motion and its small oscillations (as well
as some other annulons to be discussed below.)

We will leave the toy model at this point, and return to gauge
theory; but clearly this toy model will be a useful tool for obtaining additional
physical insights into interactions, solitons, scaling laws, decay rates, etc.
Some simple examples are given in our concluding section.

\subsection{The hadrons in the KS case}

The Hamiltonian \eref{TheHamiltonian} has a natural interpretation as
describing a sector of the hadronic spectrum of the gauge theory.  We
will first discuss this in the context of the KS theory, which is
easiest to interpret.

\subsubsection{The constituents in the KS theory}

To understand the hadrons in question, we need to understand the
various objects that carry charge in the gauge theory.  The
massless fields of the gauge theory are those of pure \none\
$SU(M)$ Yang-Mills; these are neutral under all anomaly-free
$U(1)$ symmetries.  However, as discussed in the appendix of
\cite{ks}, there are massive fields left over from the duality
cascade.  There are four chiral supermultiplets in the adjoint
representation of $SU(M)$, charged as $({\bf 2},{\bf 2})$ under
$SU(2)_\ell\times SU(2)_r$.

These emerge in the following way.  The second to last stage of the cascade
involves the gauge group $SU(2M)\times SU(M)$, with fields $A_1,
A_2$ in the $({\bf 2M},{\bf \overline M})$ representation, and
fields $B_1,B_2$ in the conjugate representation; these fields
are doublets under $SU(2)_\ell$ and $SU(2)_r$ respectively.  The
gauge group $SU(2M)$ has $2M$ flavors, so it confines
\cite{Seibergexact}; in this process
$SU(M)$ is mainly a spectator to the $SU(2M)$ dynamics. Among the
resulting bound states are the four fields
\be
(N_{ij})^{\alpha}_\beta=(A_i)^\alpha_a(B_j)^a_\beta -
{1\over M}\delta^\alpha_\beta{\rm tr}(A_iB_j)
\ee
which are in the {\it adjoint} representation of the spectator
$SU(M)$ group, with indices $\alpha,\beta$; indices $a$ are in the confining
$SU(2M)$ group.  The superpotential
\be W \propto {\rm tr}\
\left( A_iB_jA_kB_\el\right) \ \epsilon^{ik}\epsilon^{j\ell}\ee
generates, after confinement, a mass term
\be W \propto {\rm tr}(N_{11}N_{22}-N_{21}N_{12}) \ee
for the $N_{ij}$; the physical mass of the $N_{ij}$, as we will see, is
of order $m_0$. We can then make
hadrons out of these heavy fields.  (Note
that there are also fields ${\rm tr}(A_iB_j)$ which are {\it singlets} of
$SU(M)$;
these ordinary mesons will play no role in the hadrons we are about
to discuss.) Since the mass term marries $N_{11}$ and $N_{22}$,
we cannot distinguish between $N_{11}$ and $N_{22}^\dagger$. (In the same
way, and for the same
reason, we cannot distinguish right-handed bottom quarks from left-handed
ones by their gauge and global quantum numbers.\footnote{Of course,
the $N_{ij}$ are not the only bound states from the $SU(2M)$ process,
or indeed from the multiple steps of the duality cascade of KS.
However, they are likely
to be the only light stable multiplets, as
would pions be in the absence of the electroweak interactions.
The other bound states are also
in the adjoint and in other representations neutral under the
center of $SU(M)$; as such they have a marginal impact on the
annulons.}

 The vacuum described by
the KS solution has the property that only the $\ZZ_{2M}$ chiral symmetry
is broken, so the $SU(2)_\ell\times SU(2)_r$ is still realized.  However,
the
geodesic that we choose is generated by the $U(1) = T^3_\ell+T^3_r$ in the
diagonal
$SU(2)$ subgroup, which we have called $U(1)_D$.
 The field $N_{11}$ carries charge 1 under
the $U(1)_D$; $N_{22}$ carries charge $-1$, and $N_{12}$ carries charge 0.
Under the other symmetry left unbroken by the choice of
geodesic, namely
$U(1)_{A}=T^3_\ell-T^3_r$,
$N_{11}$ and $N_{22}$ carry charge 0, $N_{12}$ carries charge 1
and $N_{21}$ charge $-1$.

\subsubsection{The lowest-lying annulon of charge $J$}

Consider the lowest-lying hadron of large charge $J$ built from $J$ of
the constituents $N_{11}$, {\it i.e.} the state of lowest energy
created by applying the operator
${\rm tr}([N_{11}]^J)$
to the true $J=0$ vacuum $\ket{\Omega}$ of the gauge theory
\be
{\rm tr}[(N_{11})^J]\ket{\Omega} \ .
\ee
 This is the natural
candidate for the vacuum $\ket{0}$
of our string theory Hamiltonian
\eref{TheHamiltonian}.  As we have seen, $-P^-=H = E - m_0J = 0$ in
the vacuum, where $E$ is the eigenvalue of $i\partial_t$, the usual
Minkowski Hamiltonian.  Our vacuum state does therefore represent a
state in the gauge theory with energy $M_0\equiv m_0 J$, and as it has
no other quantum numbers or degrees of freedom, it is natural to
interpret it as the lowest-lying spin-zero hadron of charge $J$.  As
such it will contain a minimal number of constituents, namely $J$ of
the heavy scalar $N_{11}$ particles and nothing else (except some
ambient superglue, formed from the masless fields of the \none\
$SU(M)$ SYM theory.)

We identify the mass of each $N_{11}$, in the mean field of all the
others, as $m_0$.  Because of collective effects among the particles,
$m_0$ need not be the same as the mass appearing in the superpotential
(which is holomorphic) or even the physical nonholomorphic mass given
by canonically normalizing $N_{ij}$ in the effective Lagrangian for
the gauge theory.  Only from string theory do we learn that the
average mass per $N_{11}$ is of the same order as glueball masses in
the gauge theory, namely $m_0$.  From the gauge theory this is a
highly nonperturbative result.  We know of no way to derive it, and
indeed it may not be true at small 't Hooft coupling.

\subsubsection{The annulon in linear nonrelativistic motion}

Of course this hadron can move, and we should be able to write its
kinetic energy.  The first term in the Hamiltonian represents its
nonrelativistic motion
\be
{{\mathcal P}_i^2\over2 m_0 J} = {|\vec {\mathcal P}|^2\over 2
M_0} \ .
\ee
We should not be surprised that we obtain only the nonrelativistic
kinetic energy; we will only be considering energies which are
parametricaly smaller than $J$, so the kinetic energy will
generally be much less than the mass $M_0$.  We see the three
worldsheet fields $x^i$ are required to be massless so that the
spatial momenta of the hadron can be correctly represented.  This
feature is presumably generic; it shows the above string theory
represents a compactification of string theory down to three
non-compact spatial dimensions.

\subsubsection{Ripples on the annulon}

That this is really a four-dimensional {\it string} (or, more
precisely, a five-dimensional string compactified and viewed under
dimensional reduction to four dimensions) is
indicated by the $\N_R+ \N_L$ term in the Hamiltonian.  Since there
are three noncompact spatial directions, the hadron will have stringy
excitations in these directions which should be controlled by the
oscillator modes on the worldsheet in the usual way.  This is clearly
the nature of this term.

Note that the spacing between the modes is {\it not} equal to the
square root of the tension $T_s$ of the confining flux tube of
the gauge theory, $\sqrt{T_s}\sim\sqrt{gM}m_0$, times the square
root of the oscillator level $N$.  Instead we find ${gM}m_0
N/J$.  From the form of the term in the Hamiltonian, it is
natural to interpret this as tension $T_s\sim gMm_0^2$ divided by
the mass of the hadron $M_0=m_0 J$.  This form is precisely what
emerges in the above toy model and justifies our interpreting
these hadrons as annulons, taking not only the mathematical form
but also the physical shape of a heavy string.

\subsubsection{Insertion of constituents controlled by symmetries}

We can guess one more of the terms in the bosonic Hamiltonian on simple
grounds.  We know there is
a hadron in the gauge theory which is the lowest lying state
created by applying
\be {\rm tr}([N_{11}]^{J+1})\ket{\Omega} \
\ee
where again $\ket{\Omega}$
is the vacuum of the gauge theory (not our ground-state annulon!)
This differs from the lowest-lying annulon of charge $J$ only through the
replacement $J\to J+1$, up to possible $1/J$ corrections which are small
at large $J$.  In particular, we know this hadron has mass $m_0(J+1)$ for
large $J$.

Now, an $SU(2)_\ell$ rotation of this state can convert it to a hadron
{\it
of equal mass} created by applying
\be\label{oneinsert}
{\rm tr}([N_{11}]^{J}N_{21})\ket{\Omega} \
\ee
to the vacuum.  This state,
which carries $U(1)_D$ charge $J$ and $U(1)_{A}$ charge $-1$,
and differs in mass from our ground-state annulon by {\it exactly} $m_0$,
should appear in our string theory.

Of course this is also true for $SU(2)_r$, which gives us an
annulon with an inserted $N_{12}$.  Can we insert an $N_{22}$ particle
by acting first with $SU(2)_\ell$ and next with $SU(2)_r$?  We can see the
answer is no from two points of view.  First, suppose
we do act with the two $SU(2)$
symmetries in succession. Starting with an annulon with $J+2$ $N_{11}$
constituents, the action of $SU(2)_\ell$ gives us an annulon with
one $N_{21}$ constituent, as in \eref{oneinsert} above.  The action of
$SU(2)_r$ then gives us a hadron of the form
\be\label{twoinserts}
{1\over \sqrt{J}}\left[\sum_{k=1}^{J-1}
{\rm tr}([N_{11}]^{k}N_{12}N_{11}^{J-k}N_{21})\ket{\Omega} \  + \
{\rm tr}(N_{11}^{J+1}N_{22})\ket{\Omega} \right]
\ee
Thus we obtain a state which predominantly has two new constituents,
one each of $N_{12}$ and $N_{21}$.  We see that the symmetries which relate
the individual $N_{ij}$ to one another act
rather differently on hadrons that already
contain large numbers of $N_{11}$ particles.

Alternatively, suppose we add an $N_{22}$ particle into an annulon
by hand; what happens to
\be\label{macroinsert}
{\rm tr}(N_{11}^{J+1}N_{22})\ket{\Omega} \ ?
\ee
We claim there is no stable hadron which has a large overlap with
this vector in the Hilbert space.  The reason is dynamical.  Although
an individual $N_{22}$ particle is stable, in the presence of many
$N_{11}$ particles it can easily convert via
$N_{11}N_{22}\to N_{12}N_{21}$.  The underlying process involves
the term $|\partial W/\partial A_1|^2 = |B_1A_2B_2|^2$ term in the Lagrangian, which allows $B_1A_2B_2 \to B_2A_2B_1$ with $A_1$ as a spectator.
(Alternatively this process can occur as $N_{22}N_{11}\to N_{12}N_{21}$,
using the $|A_2B_2A_1|^2$ interaction.)  Once this conversion takes place,
the $N_{12}$ and $N_{21}$ can separate from one another within the annulon,
and the chance of them recombining into an $N_{22}$ is phase-space
suppressed –-- clearly of order $1/J$.

Similarly, if we act multiple times with $SU(2)_\ell$ and/or $SU(2)_r$
(adding $N_{11}$ particles so that
their number, and the $U(1)_D$ charge, remain
equal to $J$,) we obtain hadrons with
arbitrary numbers of $N_{12}$ and $N_{21}$ particles
inserted, but no $N_{22}$ particles, as long as the number of inserted
constituents is small compared to $J$.    From the symmetry
arguments we know the masses of these hadrons
differ from our vacuum annulon by
integer units of $m_0$, that they carry integer
charges under $U(1)_A$, and that they have no string
oscillation modes excited (as they are related by symmetry
to a vacuum annulon.) Therefore we
predict the existence of two worldsheet operators with
$U(1)_A$ charge $\pm 1$ which can
insert an $N_{12}$ or $N_{21}$ particle into the annulon;
these should
be related by a symmetry, and should change the mass of the
hadron by exactly $m_0$.  This expectation is borne out, as
discussed after Eq.~\eref{TheHamiltonian}. The $v,\bar v$
world-sheet fields, which descend from the part of the $S^3$
transverse to the geodesic, have precisely the right charges.
The non-oscillatory mode associated to $v,\bar v$, applied on the
string theory vacuum, leaves the $U(1)_D$ charge unchanged but
gives $H = -P^- = m_0$, or $E = m_0 (J+1)$, with $U(1)_A$ charge
$\pm 1$, as predicted.

We notice, from this structure, that the nondiagonal
$SU(2)_\ell$ and $SU(2)_r$
generators are not operators in our string theory, because they carry
nonzero $P^+$ --- they are charged under $U(1)_D$.    Rather, they are
operators which connect the Hilbert space of
the string theory of charge $P^+=J$ to that
with charge $P^+=J\pm 1$.  In this way these symmetries can remain exact
in the full theory but be absent within any one charge sector.

\subsubsection{Supersymmetry}

  Since the gauge theory is \none\ supersymmetric, we
expect that the ground-state annulon (and indeed every state in
the Hilbert space of the bosonic Hamiltonian) has a fermionic
superpartner, a massive fermion, with the same mass and charge
(and thus the same $P^-$.)  Indeed, as the annulon is massive and
charged, its multiplet structure is that of a complex multiplet,
in particular the combination of a chiral multiplet of charge $J$
and a chiral multiplet of charge $-J$.  This implies two complex
scalar fields and a Dirac fermion (the same as the
electron-selectron supermultiplet in SQED.)  In a dual string
picture, the charge $J$ components of the complex multiplet can be
generated using the zero modes of two massless worldsheet
fermions on a state of fixed $p^+$. Taking into account the
left-moving and right-moving contributions to the closed string
Hilbert space, these zero-modes form two raising and two lowering
operators generating a four-dimensional Hilbert space: two
bosonic states and two fermionic. (Note our vacuum annulon is a
complex boson.) As noted in section 5.2, two is precisely the
number that we have.  The action of one raising zero mode
converts an $N_{11}$ constituent to its $\psi_{11}^\alpha$
fermionic partner ($\alpha$ a spin index), and thus
\be\label{onesusyacts}
{\rm tr}[N_{11}^J]\ket{\Omega}  \  \Rightarrow \
{\rm tr}[N_{11}^{J-1}\psi_{11}^\alpha]\ket{\Omega}.
\ee

Since we have the massless zero modes corresponding to the
insertion of $\psi_{11}$,
and since we have bosonic modes of mass $m_0$ corresponding to the insertion
of $N_{12}$, why do we not have a fermionic mode of mass $m_0$
associated to $\psi_{12}$?  The absence of this operator follows
from the same logic \eref{oneinsert}-\eref{macroinsert} that
explains the absence of a mode for $N_{22}$.
We might expect that the combination of supersymmetry and $SU(2)_r$
would turn $N_{11}\to \psi_{12}$, but in a hadron with $J$ constituents
the action of these two symmetries instead preferentially inserts one
$N_{12}$ constituent and one $\psi_{11}$ constituent.  Similarly,
the transition $\psi_{12}N_{11}\to N_{12}\psi_{11}$ can be mediated
by $SU(2M)$-gluino exchange among the $A_i, B_j$ particles and their
fermionic partners; as before, once this transition occurs the $\psi_{12}$
particle is unlikely to be reconstituted.

The same logic applies to the action of two supersymmetries; rather
than generating the highest component of the $N_{11}$ chiral supermultiplet,
the double action of supersymmetry inserts two $\psi_{11}$ constituents
into the annulon.

\subsubsection{The non-universal directions}

The remaining structure of the Hamiltonian --- the $z$ and
$u,\bar u$ zero modes and the excited states in the five internal
directions --- cannot be predicted by any symmetry arguments. The
only additional feature determined by symmetry is that in KS the
transformation $SU(2)_\ell\leftrightarrow SU(2)_r$ which
exchanges $v$ and $\bar v$ should be accompanied by
$u\leftrightarrow -\bar u$ and $z\leftrightarrow -z$; but this
symmetry is absent in MN. For this reason we expect the remaining
features of the string theory to differ from model to model, and
indeed the MN and KS cases are different.

We have made a number of attempts to interpret the non-universal
directions in the KS theory.  There are multiple possibilities,
motivated by a variety of different arguments.  However, we have been
unable to determine which of these possible interpretations is
correct, if any.  We leave this issue for further research.

One important additional comment is that we have been a bit
cavalier in specifying our hadrons.  While it is true that our
ground state hadron is the lowest-lying hadron created by ${\rm
tr}(N_{11}^J)$ acting on the vacuum, it is not true that the
hadron contains only $N_{11}$ particles.  It may also contain
$N_{22}^\dagger$ particles, and its wave function involves some
mixture of possible states.  (In a similar way, a proton may be
created by $u_Lu_Ld_L$ or by $u_Ru_Rd_R$; the true proton has a
wave function rather different from that suggested by either of
these operators.)  We have not determined the wave function of
the ground state hadron unambiguously, and an understanding of
the non-universal directions may require further investigation of
this issue.  Certainly there need not be any simple relation
between the {\it operators} associated with the conformal
conifold-derived pp-wave and the {\it states} associated with our
pp-wave for the confining KS theory.

\subsubsection{Summary}

We now largely understand what our string theory is describing.
The vacuum is a long, stable annulon of massive $N_{11}$ (and
$N_{22}^\dagger$) particles. This heavy object can move in
rectilinear non-relativistic motion, and it can wriggle as a
non-relativistic string.  We can also insert into the chain of
these particles any number of $N_{12}$ and $N_{21}$ particles. The
original and inserted objects have a probability amplitude for
their locations on the annulon, described by a wave function.  The
various energy eigenstates for this wave function give various
hadronic states which correspond to strings with various
excitations in the massive directions. The zero modes of the
massive directions correspond to inserting the fields $N_{ij}$,
and their conjugates, with constant amplitude around the string.
This is of course consistent with BMN, but differs from it just as
one would expect a description of states to differ from a
description of operators.\footnote{ The structure of
\eref{TheHamiltonian} also hints at a possible analogue of the
non-relativistic quark model.  In such a model, the stringy
excitations in the massless directions could be interpreted as the
insertion of ``constituent'' gluons into the ground-state annulon.
For the massive directions, the excitations would simply involve
insertions of $N_{12}$, $N_{21}$, {\it etc.} We leave this idea
for future investigation.}

\subsection{More on the MN case}

Now let us turn to the MN case.  We have already seen that the KS case
has a geodesic direction, two $v$ directions, two $u$ directions, a
$z$ direction and three $x^i$ directions.  The $x^i$ directions are
massless, from translational symmetry in real space, and the $v$'s
have mass $m_0$, from the $SU(2)$ symmetries which rotate the geodesic
into the rest of the $S^3$.  Also, there are two massless fermionic
zero modes, by supersymmetry.  The MN theory must share all of these
features, which follow simply from symmetries, and as discussed in
section 5, it does, though the supersymmetry is somewhat obscured.

 The MN case involves the dimensional reduction of the
six-dimensional \ntwo\ Yang-Mills theory on a two-sphere, with the
appropriate twisting to maintain \none\ supersymmetry in four
dimensions.  The massless six--dimensional \ntwo\ vector multiplet
can be split into a six--dimensional \none\ vector multiplet and a
six--dimensional \none\ hypermultiplet; the supercharge which
survives the two--sphere reduction is in this \none\ sub--algebra.
After reduction, the massless six-dimensional \none\ vector
multiplet gives a massless four-dimensional \none\ vector
multiplet, along with a tower of massive Kaluza-Klein \none\
vector multiplets in the adjoint representation.   The lowest of
these are three massive \none\ vector multiplets in a triplet of
$SU(2)_r$.

   The other part of the six--dimensional \ntwo\ vector
multiplet, the six--dimensional \none\ hypermultiplet, gives only
massive Kaluza-Klein modes after reduction on the two--sphere.
These are all in four--dimensional complex \none\ chiral
multiplets and will transform under both $SU(2)$ symmetries.  The
lightest of these modes form two massive complex \none\ chiral
multiplets (eight real scalars) which transform in the
bi--fundamental of $SU(2)_\ell\times SU(2)_r$. Thus both the KS
and MN gauge theories have scalars corresponding to motion on the
three-sphere at the minimum $AdS$ radius.  In the MN case these
motions are generated by the (now twisted) R-symmetry of the
six-dimensional theory, while in KS they are the angular modes
which rotate the $N_{11}$ particles into $N_{12}$ and $N_{21}$.
The symmetries of the three-sphere are enough to predict that the
properties of the $v,\bar v$ world-sheet fields in the MN case
are the same as they are in the KS example.

 For those phenomena not controlled by symmetries, the
theories may, and do, differ.  In particular, they differ on the
masses $\hat m_z$ and $\hat m_u$, even when we account for the
shift in mode frequencies discussed in section 5.2, which is
needed to see the fermionic zero modes and supersymmetry.  As
discussed at the end of section 3, the MN case lacks the
left-right symmetry of the KS case. Meanwhile the KS case has a
surviving nonzero NS-NS 2-form, which the MN case lacks. All of
these effects contribute to the differences between the two
string theories.

The MN case does pose an additional problem: the $z$ direction is
massless. Note, however, that this does {\it not} necessarily mean
that it corresponds to a massless constituent, nor do we need to treat
it as we treat the massless $x^i$ directions. Rather, it simply means
that there is an excitation internal to an MN annulon which costs
energy minus charge (both of which could be nonzero) much less than
$m_0$ as $J\to \infty$.  At the very least we expect this apparent
flat direction, which clearly is not present in the full MN metric, to
be lifted at finite $J$; this is in contrast to the $x^i$
directions which are massless by translational symmetry.  This issue
deserves further investigation however.

  With the exception of this issue, the KS and MN theories
are qualitatively similar in most respects and quantitatively
equal where they ought to be.  It would be interesting to compare
these theories to the \none* case, if the latter is tractable.

\section{Wilson loops with charge}

In this section we return to the question of confinement and
states of large $J$. We should be able to see that the theory is
confining by examining Wilson loops.  In the dual ten-dimensional
supergravity, a straight tube of electric flux appears as a
fundamental string extended in one of the Minkowski spatial
directions and placed at a definite $AdS$ radius $r$ (and at a
definite point in the other coordinates as well.)  Such a string
will fall to smaller $r$ unless dynamically prevented from doing
so.  In a conformal theory it can fall to $r=0$, where its
tension, proportional to $\sqrt{g_{tt}g_{xx}}$, goes to zero
\cite{wl}. In a confining gauge theory there is a minimum value of
$\sqrt{g_{tt}g_{xx}}$, at some radius $r_{0}$.  The string
plummets to this radius but can go no further; its tension is
bounded from below \cite{confinewl,LS}.
A heavy quark-antiquark pair at
$\vec {\bf x}= (0,0,0)$ and $\vec{\bf y}= (\ell,0,0)$ in the
gauge theory, and the chromoelectric flux between them,
correspond to a single string which ends on the boundary
$r\to\infty$ of the gravity background at the points $\vec{\bf
x}$ and $\vec{\bf y}$ \cite{wl}.  The energy $V(\ell)$ of the
configuration in the gauge theory is proportional to the total
energy of the string (note it is formally infinite because the
quark masses are infinite, but $dV/d\ell$ is finite.)  If
$dV/d\ell$ contains an $\ell$-independent additive constant $T$,
then $V$ contains a term equal to $T\ell$, and the theory has
linear confinement, with a flux tube of constant tension $T$.
This happens precisely when the string with its ends fixed at
$\vec{\bf x}$ and $\vec{\bf y}$ falls down to $r_{0}$ but can
fall no further, and lies there like a rope resting in a
flat-bottomed lake.  Since the string lies entirely at $r_{0}$, we
might expect to detect its tension by looking at geodesics near
$r_{0}$.

\begin{figure}[th]
\label{drapedstring}
\begin{center}
 \centerline{\psfig{figure=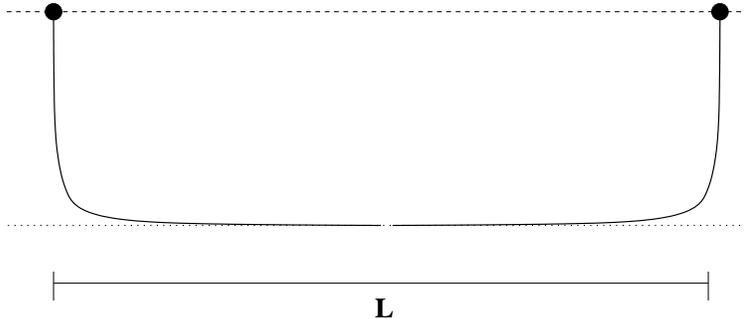,width=10cm,clip=}}
 \caption{In the string dual of a confining theory, the quark and
 antiquark sources are given by the $r$-dependent segments near
 the ends of the string, while the flux tube between them is given
 by the region of the string lying at $r=r_0$. }
\end{center}
\end{figure}

But flux tubes of this sort know nothing about global charges,
whereas we know that plane wave backgrounds involve
considerations of highly-charged systems.  How should we connect
the two? One might think that one should consider quarks carrying
global charge, but this is not correct.  Instead, one should
consider a quark-flux-antiquark {\it system} carrying large
global charge.  Such systems are not familiar from QCD, for a
variety of reasons.  However, in our theory, they are easy to
construct, because we have massive adjoint
fermions $\psi$ and scalars $\phi$ carrying a global $U(1)$
charge, with mass $m_0$.  To such a gauge theory, we may add very
heavy quarks $q$ ($m_q\gg m_0$) as probes of the system.  We can
then form heavy hadrons of the form $\bar q q$, $\bar q\phi q$,
$\bar q\phi\phi q$, {\it etc.}, of charge 0,1,2, {\it etc.}
Starting with a hadron $\bar q(\phi)^J q$, we can form a {\it
string} labelled by $J$, its total charge --- {\it not} its
charge density!  --- by pulling the $q$ and $\bar q$ in the
hadron until they are a distance $\ell$ apart. Our system is then
made from a $q$, a $\bar q$, some gluons, $(n+J)\phi$ particles
and $n$ $\bar \phi$ particles, where $n$ may be fluctuating but
$J$ is constant.  The lowest energy eigenstate of this system has
energy $V(\ell;J)$, a function one might attempt to calculate.
For $\ell$ very large and $J$ held finite, the effect of the $J$
widely-scattered $\phi$ particles will be negligible, and one will
find $V(\ell,J)\approx 2 m_q+T\ell$. Conversely, for finite
$\ell$ but with $J$ taken very large, one will find $V(\ell;J)$
of order $2 m_q + Jm_0$. Certainly, then, by looking at null
geodesics at $r=r_{0}$, and looking at large $J$ but even larger
$\ell$, we can detect whether there are confining flux tubes in
the theory.  But we will not have to work so hard.  Instead,
working at large $J$ and looking at the leading $\ell/J$
correction to $V(\ell,J)$, we will show the tension $T$ is
nonzero.

In our theory, we can identify the $\phi$ particles with
$N_{11}$.  If $J$ is very large, so that the $N_{11}$  particles are
very densely distributed on the flux tube, then we would expect
these globally-charged flux tubes to be made from
annulonic material.  We would naturally guess, from our string theory,
that just as excited states of the lowest-lying annulon
have energy {\it quadratic} in the string tension, so $V(\ell;J)$
will be quadratic in $T$ for very large $J$.

In fact, a semiclassical field theory analysis can already suggest for
us the form that we should find.  Let $T$ be the tension of a
confining string with $J=0$.  If the mass of the adjoint particles is
$m_0$, we might guess the mass of a string of length $\ell$ would be
$T\ell+Jm_0$.  But this is not correct; it does not account for the
binding of the particles to the string.  Instead, we must treat the
the global charge that the string can carry in much the way one treats
the electric charge of dyons --- using collective coordinates.

We may get insight by taking a partly S-dual version of this
problem, in which we bind heavy particles of mass $m_0$ with a
conserved global charge to a magnetic flux tube.  A still easier
version is given by compactifying such a problem on a circle, so
that the magnetic flux tubes themselves become particles
(vortices in 2+1 dimensions) and the binding problem becomes
familiar.  The vortices are solitons and the global charge they
carry should be treated as in any collective coordinate problem,
leading semiclassically to a formula for a soliton of vortex
charge $p$ and global charge $q$:
\be
M_{p,q} = \sqrt{p^2 m_{vortex}^2 + q^2 m_0^2}
\ee
just as with dyons in four dimensions.  Lifting the
problem back to four dimensions we replace $m_{vortex}$, 
with $TR$ (where $R$ is the radius of the compact dimension.)
Indeed this can be seen in string theory, where one could
consider various string-brane semiclassical bound state formulas, which
are always governed by quadratic Born-Infeld formulas.

Following this logic, one is led to suspect that a similar
formula governs the binding of heavy particles to an electric
flux tube
\be\label{dyonlike}
V(\ell;J) = \sqrt{T^2 \ell^2 + J^2 m_0^2}.
\ee
This of course matches our naive expectations in the large $\ell$
and large $J$ limits. And for large $J$, we have
\be
V(\ell;J) \approx J m_0 + {1\over 2} {T^2 \ell^2 \over J m_0},
\ee
another ``non-relativistic'' formula.  From this we learn that
we can detect confinement in the gauge theory by looking
at the order-$\ell^2$ term --- not at an order-$\ell$ term! --- in $V(\ell;J)$.
In general this formula might be subject to nonlinear corrections,
but since we are in a limit where strings behave
classically, we would not
be surprised if such corrections were absent.

Of course this is also what is obtained in our toy model.  Suppose we take a
string on $M^8\times S^1\times S^1$, where the radii of the circles
are $R_1$ and $R_2$, and we wrap the string on one circle while
boosting it in the other.  Before boosting, the string has mass
$R_1/\alpha'$.  When boosted by $J$ units of KK momentum ({\it
not} momentum per unit length) the
string appears, from the nine-dimensional point of view, to be a
static string of mass
\be \sqrt{\left({R_1\over \alpha'}\right)^2 + \left({J\over
R_2}\right)^2} \ee
which matches the formula above, if we identify $1/R_2 = m_0$,
$1/\alpha'=T$ and $R_1=\ell$. Again, at large $J$, we see the
first correction to $J/R_2$ is {\it quadratic} in $R_1$.

Finally, we now calculate this in the supergravity dual to the
gauge theory, by generalizing the results of \cite{wl}.  In a
confining theory, the energy of a system consisting of a heavy
quark and a heavy antiquark at a distance $\ell$ from one
another, and with no global charges, is given in supergravity by
the total energy of a semiclassical string whose endpoints
contact the boundary at spatial positions $\vec {\bf x}=(0,0,0)$
and $\vec {\bf y}= (\ell,0,0)$.  When $\ell$ is very large, and
the theory is confining, the string becomes very simple, as shown
in figure 1.  At the two ends, the string descends rapidly from
the $AdS$ boundary to the radius $r_0$ where the tension of the
string is minimized.  This behavior is $\ell$-independent and the
energy of these two regions correspond to the constant masses of
the heavy quark and antiquark.  The majority of the string lies
along a line from $\vec {\bf x}=(0,0,0)$ to $\vec {\bf y}=
(\ell,0,0)$ but lying at $r=r_0$.  The tension of the string in
this region is constant, so the energy of this part of the string
grows linearly with $\ell$; the constant of proportionality gives
the tension of a confining string in the gauge theory.

Our claim is that the addition of global charge to this system is as
simple as taking the above configuration and giving it a definite
momentum along one of the $S^1$ directions in the compact
five-dimensional space.  We will now show that we reproduce the above
expectations for $V(\ell;J)$.

We start by considering the Nambu-Goto action:
\be
S={1\over 2\pi \a'}\int d\tau d\sigma \sqrt{-\mbox{det}\,G_{MN}\pd_\a X^M
  \pd_\b X^N}.
\ee
The general form of the classical configuration we are interested in
involves an open string with its ends at two points on the boundary of
the bulk space.  Its
radial position varies with $\sigma$, and it has in general
nontrivial motion along an angle $\phi$.  In short, the string has a worldsheet
of the form
\be
t=t(\tau), \quad x=x(\sigma), \quad \p=\p(\tau), \quad r=r(\s).
\ee
The relevant part of the metric is therefore,
\be
ds^2 =-g_{tt}dt^2 + g_{xx}dx^2 + g_{rr}dr^2 + g_{\p\p}d\p^2 + \ldots
\ee
Evaluating the Nambu-Goto action on this background we obtain
\be
S={1\over 2\pi \a'}\int d\tau d\s  \sqrt{(g_{tt}\dot{t}^2-
g_{\p\p}\dot{\phi}^2)(g_{xx}+g_{rr}(\pd_\s r)^2)}.
\ee
The energy and angular momentum (global charge), as conjugate variables
to $t$ and $\phi$, are
\bea
E&=&{1\over 2\pi \a'}\int d\s\,  g_{tt}\,\,\dot{t}
\left({g_{xx}(\pd_\s x)^2 + g_{rr}(\pd_\s r)^2)
\over g_{tt}\dot{t}^2- g_{\p\p}\dot{\phi}^2}\right)^{1/2}, \nonumber\\
J&=&{1\over 2\pi \a'}\int d\s\,  g_{\phi\phi}\,\,\dot{\phi}
\left({g_{xx}(\pd_\s x)^2 + g_{rr}(\pd_\s r)^2)
\over
g_{tt}\dot{t}^2- g_{\p\p}\dot{\phi}^2}\right)^{1/2}.
\eea

The precise minimization of the energy subject to fixed charge $J$ would
be quite involved in the backgrounds
we consider. For example, for the KS background
the warp factor is not known analytically for all values of the
radius. However, using a natural ansatz, we can obtain an excellent
estimate for the relationship between the energy and
the angular momentum, and later justify how
any contribution left unaccounted for is appropriately suppressed.

We naturally fix the static gauge
$t=\tau$ and $x=\ell{\s\over 2\pi}$,
where $\ell$ is the total length of
the string (as measured in the gauge theory, using the Minkowski
metric!).  Most of the string
lies along the ``minimal'' radius $r_0$
\be
r[x(\s)] = r_0 , \ \delta < x < \ell-\delta
\ee
and $r[x]\to\infty$ as $x\to 0$ and as $x\to \ell$.  We assume
that $\delta$ is independent of $\ell$ for large $\ell$, which
corresponds to the physical expectation that the quark and
antiquark sources do not grow as $\ell$ increases. We also choose
that the endpoints do not rotate, so that the quark and antiquark
sources do not themselves carry any global charge. This means
that $\dot\phi\to 0$ at the ends of the string at $\s=0,2\pi$.
Finally --- and this is the least obvious part of the ansatz ---
we assume that the majority of the angular momentum is spread
{\it uniformly} in the region far from the ends, by taking the
majority of the string to move uniformly on a circle
parameterized by $\phi$:
\be
\phi \approx \omega\tau, \ \delta < x < \ell-\delta.
 \ee
Altogether this implies
\be
J\approx {\omega\over 2\pi \a'} \,\,
\sqrt{{g_{\phi\phi}^2g_{xx}(\ell/2\pi)^2 \over g_{tt}-
g_{\p\p}\omega^2}\Bigg|_{r=r_{0}}}
\int_{2\pi\delta/\ell}^{2\pi(1-\delta/\ell)} d\s \approx {\omega
\ell\over 2\pi\a'} \,\, \sqrt{{g_{\phi\phi}^2 g_{xx} \over
g_{tt}- g_{\p\p}\omega^2}\Bigg|_{r=r_{0}}}.
 \,
\ee

With this gauge and ansatz we have
\bea
E&\approx &{1\over 2\pi \a'}\int_0^{2\pi\delta/\ell} d\s\,
g_{tt}\,\, \left({g_{xx}(\ell/2\pi)^2 + g_{rr}(\pd_\s r)^2)
\over g_{tt}- g_{\p\p}\dot{\phi}^2}\right)^{1/2} \nonumber\\
&&+ {1\over 2\pi \a'}\int_{2\pi(1-\delta/\ell)}^{2\pi} d\s\,
g_{tt}\,\, \left({g_{xx}(\ell/2\pi)^2 + g_{rr}(\pd_\s r)^2)
\over g_{tt}- g_{\p\p}\dot{\phi}^2}\right)^{1/2} \nonumber\\
&&+ {1\over 2\pi \a'} \,\,
\left({g_{tt}g_{xx}(\ell/2\pi)^2
\over g_{tt}- g_{\p\p}\omega^2}\Bigg|_{r=r_{0}}\right)^{1/2}
\int_{2\pi\delta/\ell}^{2\pi(1-\delta/\ell)} d\s\,   \nonumber\\ \nonumber\\
&\approx& 2{\bf m}_q + {g_{tt}\over g_{\p\p}}\Bigg|_{r=r_{0}}{J\over\omega}
\eea
where
\be
{\bf m}_q \equiv {1\over 2\pi \a'}\int_0^{2\pi\delta/\ell} d\s\,
g_{tt}\,\, \left({g_{xx}(\ell/2\pi)^2 + g_{rr}(\pd_\s r)^2) \over
g_{tt}- g_{\p\p}\dot{\phi}^2}\right)^{1/2} \approx {1\over 2\pi
\a'}\int_0^{2\pi\delta/\ell} d\s\,\, \sqrt{ g_{tt}g_{rr}}(\pd_\s
r)
 \ .
\ee
This last expression is divergent, representing the infinite mass of
the heavy quark, but more importantly for our purposes it is
essentially $\ell$-independent, and gives a physically inconsequential
additive shift to the energy $V(\ell;J)$.

More succinctly, defining
\be
m_0 \equiv
\lim_{r\to r_0}
\sqrt{g_{tt}/g_{\phi\phi}}
\ee
as we did both for KS and MN, we have
\be
J\approx {g_{tt}\ell\over 2\pi \a'}
 {\o/ m_0^2\over \sqrt{1 - \o^2 /m_0^2}} \ , \
E\approx 2{\bf m}_q + {g_{tt}\ell\over 2\pi \a'}
{1\over \sqrt{1 -  \o^2 /m_0^2}} \ .
\ee
whence
\be
V(\ell;J) =E \approx 2{\bf m}_q + \sqrt{ {g_{tt}^2\, \ell^2\over
    (2\pi\,\a')^2} +  m_0^2\,J^2 }.
\ee
Recalling that
the tension is $T=\sqrt{|g_{tt}g_{xx}|}/(2\pi\a')= g_{tt}/2\pi\a'$, we reproduce the
formula \eref{dyonlike}
obtained by the field theory analysis:
\be
V(\ell;J)\approx \sqrt{T^2\, \ell^2 + m_0^2\,J^2} + {\rm constant}.
\ee
In an appendix, we show that our ansatz is stable and that all corrections
are of order $1/\ell$ or $1/J$ relative to the terms that we have kept.

\section{Closing Comments}

We have found a sector of a gauge theory whose hadrons are described
by an exactly soluble string theory.  We obtained them through a
Penrose limit around a geodesic sitting at the minimum $AdS$ radius,
moving in Minkowski time, and winding around a circle on the compact
part of the bulk space.  We obtained a description of hadrons of
charge $J$ with mass of order $J$, which we argued were of the form of
nonrelativistic strings.  The string theory describes their motion,
their ripples, their superpartners and their global symmetry partners.

It is important to emphasize a mathematically essential point that
makes our construction possible.  One of the key characteristics
of the Einstein equations is their nonlinearity, which implies
that the expansion in the metric around a neighborhood of a
particular point is not a well-defined procedure. For example, in
the context of the KS background, the Ricci flatness of the
six-dimensional space perpendicular to the Minkowski directions is
required in order to satisfy the equations of motion. The deformed
conifold is, of course, Ricci flat. However, truncating the metric
near the apex of the deformed conifold to second order results in
a space metrically equivalent to ${\bf R}^3 \times S^3$ which is
no longer Ricci flat and thus ceases to satisfy the equations of
motion. The use of such approximations is equivalent to neglecting
back--reaction in many situations, and although one might extract
sensible results it is not a consistent procedure in general.

In the context of the Penrose limit \cite{penrose,gueven,blau}
however, there is a well-defined expansion around a null
geodesic. Properly interpreted, this amounts to having to consider the
expansion up to second order around the null geodesic. In particular,
in the KS solution, expanding the metric around the end of the
conifold at $\tau=0$ (here $r\propto \cosh\tau$), keeping up to
quadratic terms in $\tau$, is a well-defined truncation in the Penrose
limit. This particular property of the limit makes our analysis exact.

It is interesting, and at first glance slightly puzzling, that we
have recovered particles moving in three spatial dimensions by
taking a Penrose limit.  It was argued in \cite{thebdyofBMN}
that the boundary of the pp-wave corresponding to conformal
field theories is a null line --- null in the bulk but timelike
from the coordinates of the field theory.  In short, the theory
is purely quantum mechanical, with no spatial dimensions.
Why, then, do we seem to have spatial directions in our limit?
In fact, we do not have them.  Although we have three spatial
{\it momenta}, which can take any finite values, we are working
in a limit where the annulon masses are infinite, and thus
{\it the spatial velocities are all zero.}  Thus the annulons
never actually move anywhere (at infinite $J$), and the theory remains quantum
mechanical despite the presence of continuous momenta.

This feature, and many others, is captured
by our toy model of a string boosted along a
compact circle.
This model can further be used to guess other dynamical
features of these particles.  For example, the stability of the
annulons to decay can easily be estimated.  An annulon
of charge $J$ can decay to two annulons of charge $J_1$ and
$J_2$.  The rate for this process is the same as that for a one-to-two
string decay in flat space, but greatly slowed down in the lab
frame by the time dilation associated with the boost.  Similarly,
the rate for two-to-two scattering is simply given by the Virasoro-Shapiro
amplitude and its generalizations.  For example, if annulons
of charge $J_1$ and $J_2$ scatter elastically, the amplitude is
given by a sum over $s$-channel annulon poles of charge $J_1+J_2$, or
equivalently by a sum over ordinary strings of charge $J=0$
in the $t$-channel.  Regge physics and other characteristic
features of string amplitudes will also be reproduced.

  Another initially puzzling observation is
that the original BMN string theory describes operators with $J^2 \ll N$,
and has an effective coupling constant $gN/J^2$.  But where
BMN finds this latter combination, we find $gM/J$, or, more
accurately, $(gM)^2/J^2$.  This phenomenon is a reflection
of an important property of the KS metric: the number of colors in
the far infrared, as measured by the integral of the 3-form through
the $S^3$, is $M$, but the far-infrared metric has each factor of
$M$ enhanced by an extra factor of $(gM)$.    Essentially, the
dual to ${\cal N}=4$ SYM (or to the far UV of KS) is controlled mainly
by a relation between the metric and the 5-form, but in the
far infrared this converts to a relation between the metric
and the 3-form.  In the process the extra factor of $gM$ appears.

Thus, in our string theory, the effective coupling constant
is $g[gM]M/J^2$, and the subleading effects which are
being neglected are of order $J^2/[gM]M$.  This latter point
is very important, because it shows that our analysis is
only good when $J\ll \sqrt g M \ll M$.  Consequently our
strings, despite having $J$ scaling like $M$, are still
far from being baryon-like giant gravitons (with $J\sim M$.)

 A related question involves the properties of annulons at small 't
Hooft coupling.  In this regime some aspects of the annulons are
presumably described using field theoretic perturbation theory, and
those of the charged Wilson loops via a semiperturbative treatment.
At present we do not know which aspects of our results continue to
this regime, and how the remainder are modified.  There are many
interesting issues to be considered here, not the least of which is
identifying the difference between annulons and giant-gravitonic
hadrons --- the latter being well-described using Hartree-Fock
mean-field techniques, as in Witten's description of large-$N$
baryons.

Finally, we would hope that this set of hadrons, which do not appear
in QCD, are not the only ones which can be treated in this fashion.
States of high {\it spin} and small charge, which are long
semiclassical strings far along Regge trajectories, do appear in
Yang-Mills theory, and to some degree in physical QCD.  It is these
states which appear in the original Chew-Frautschi plots of the
so-called Regge trajectories of QCD.  Any improvement in our ability
to study these states would be of substantial physical interest.

\

\begin{center}
{\large  {\bf Acknowledgments}}
\end{center}

We have benefited from discussions with D. Berenstein, S. Cherkis,
J. Figueroa-O'Farrill, J. Gomis, A. Hashimoto, C. Herzog, I.
Klebanov, J. Maldacena, A. Naqvi, C. N\`u\~nez, N. Seiberg,  C.
Thorn and A. Tseytlin.  J.S. and M.J.S. thank the Institute for
Advanced Study, where much of this work was done.  E.G.G.,
L.A.P-.Z and M.J.S are grateful to the Aspen Center for Physics
for hospitality at various stages of this work. E.G.G. is
supported by Frank and Peggy Taplin, and by NSF grant PHY-0070928.
L.A.P.-Z. is supported by a grant in aid from the Funds for
Natural Sciences at I.A.S. The work of J.S was supported in part
by the US-Israel Binational Science Foundation and by the Israel
Science Foundation.  M.J.S was supported by DOE grant
DOE-FG02-95ER40893, NSF grant PHY-0070928, and an award by the
Alfred P. Sloan Foundation.

\newpage
\appendix
\section{Deformed Conifold: Metric and Symmetries}
\subsection{New Coordinates for the Deformed Conifold}

To find a convenient set of coordinates, we go back to the basics
\cite{candelas}. The deformed conifold is defined in terms of a
complex two-by-two matrices $W$ satisfying:
\be
{\textrm {det}}\,W = -{\ep^2\over 2}.
\ee
We define the variable $\tau$ by setting:
\be
{\textrm tr}(W^\dag\ W) = \ep^2\cosh\tau
\ee
A simple solution to this equation is
\be
W_{\ep} = {\ep\over \sqrt{2}}\, \Bigg(
\begin{array}{cc}
e^{\tau\over 2} & 0  \\
0 & -e^{-\tau\over 2} \\
\end{array}\Bigg)
\;=\;{\ep\over \sqrt{2}}\,e^{{\tau\over 2}\sigma_3}\,\sigma_3,
\ee
and we can generate the whole set by acting on the left and right
with $SU(2)$ matrices $L$ and $R$:
\be
W = L\,W_{\ep}\,R^\dag.
\ee

For reasons which will soon be obvious, we choose to re--write
the two $SU(2)$ matrices $L$ and $R$ as:
\be
L = T\,S,\qquad R = \sigma_3\,S\,\sigma_3.
\ee
Now for $\tau \neq 0$ we see that if we take
\be
S \rightarrow S\, e^{i\theta\,\sigma_3},
\ee
then $W$ remains invariant.  This means that $S$ represents
coordinates for $SU(2)/U(1) = S^2$.  Also, when $\tau = 0$, $W$
takes the form
\be
W = T\,S\,W_0\,\sigma_3\,S^\dag\,\sigma_3 = T\, W_0 = T\,\sigma_3,
\ee
which means that $T \in SU(2)=S^3$ is a good coordinate for the
$S^3$ at the origin and gives a coordinate independent of $S$ for
$\tau\neq\ 0$.

We will now compute the deformed conifold metric in these new
coordinates, starting from \cite{candelas,mt}:
\be
\label{Kmetric}
ds_6^2 = \ep^{-{2\over 3}}K(\tau){\textrm tr}(dW^\dag\ dW) +
\ep^{-{8\over3}} \sinh^{-1}(\tau) K'(\tau) |{\textrm tr}(W^\dag\
dW)|^2.
\ee
We first write
\be
T^\dag\,dT = - dT^\dag\,T = {i\over 2}\,\omega^a\,\sigma_a,
 \qquad S = e^{{i\over 2}\,\phi\, \sigma_3}\,e^{-{i\over 2}\,\theta\,
 \sigma_2},
\ee
and then
\bea
\label{firstpart}
{\textrm tr}(dW^\dag\ dW) &=&
 {\ep^2\over 4}\,\cosh(\tau)\big(d\tau^2 + (\omega^a)^2\big) \nonumber \\
 &+& {\ep^2}\,\sinh^2({\tau\over 2})\;
 \bigg[\;
 (d\theta^2 + \sin^2\theta\,d\phi^2) \\
 &&- \,(\sin\phi\,\omega^1 + \cos\phi\,\omega^2)(d\theta) \nonumber\\
 &&- \,(\cos\theta\cos\phi\,\omega^1 - \cos\theta\sin\phi\,\omega^2
 - \sin\theta\,\omega^3)(\sin\theta\,d\phi) \nonumber
 \bigg]
\eea

and
\bea
\label{secondpart}
|{\textrm tr}(W^\dag\ dW)|^2 =
 {\ep^4\over 4}\,\sinh^2(\tau) \bigg[\;d\tau^2\; + &&\\
 (\sin\theta\cos\phi\,\omega^1 &+&
 \sin\theta\sin\phi\,\omega^2
 +\cos\theta\,\omega^3)^2\bigg]  \nonumber
\eea
\subsection{Connection to Other Coordinates}
We would like to connect our coordinates above with the ones used
in \cite{ks} (inherited from Minasian and Tsimpis
\cite{mt}). These coordinates can be written in the following
manner:
\bea
W &=& L_1\,W_{\ep}\,\sigma_3\,\sigma_1\,L_2^\dag, \nonumber \\
L_1 &=& e^{{i\phi_1\over 2}\,\sigma_3}\,e^{{-i\theta_1\over
2}\,\sigma_2}\,e^{{i\psi\over 4}\,\sigma_3},\, \label{oldcoordinates} \\
L_2 &=& e^{{i\phi_2\over 2}\,\sigma_3}\,e^{{-i\theta_2\over
2}\,\sigma_2}\,e^{{i\psi\over 4}\,\sigma_3}.\, \nonumber
\eea
We can now rewrite the 1--forms $g^i$ derived from these
coordinates in terms of the 1--forms which come from $T^\dag dT$
and $S^\dag dS$.  One first writes a change of basis for the
1--forms $\omega^i$,
\be
  \sqrt{2}\,\tilde{g}^3\,\sigma_1-\sqrt{2}\,\tilde{g}^4\,\sigma_2+\tilde{g}^5\,\sigma_3 \;=\;
  S^\dag\,\omega^a\,\sigma_a\,S,
\ee
which gives
\bea
  \tilde{g}^5 &=& \sin\theta\cos\phi\,\omega^1 - \sin\theta\sin\phi\,\omega^2
 + \cos\theta\,\omega^3, \nonumber\\
- \tilde{g}^4 &=& {1\over\sqrt{2}}\,(\sin\phi\,\omega^1 + \cos\phi\,\omega^2), \\
  \tilde{g}^3 &=& {1\over\sqrt{2}}\,(\cos\theta\cos\phi\,\omega^1 -
  \cos\theta\sin\phi\,\omega^2 - \sin\theta\,\omega^3). \nonumber
\eea
After a little algebraic work one finds
\bea
  e^{-{i\over 2}\psi}\,(g^3 + i\,g^4) &=&  \tilde{g}^3 + i\,\tilde{g}^4   \nonumber \\
  g^5 &=& \tilde{g}^5  \\
  e^{-{i\over 2}\psi}\,(g^1 + i\,g^2) &=&  (\tilde{g}^3 -
  \sqrt{2}\sin\theta\,d\phi) + i\,(\tilde{g}^4 +
  \sqrt{2}\,d\theta) \nonumber
\eea
Now we can rewrite the expression (\ref{firstpart}) and
(\ref{secondpart}) as
\bea
\label{rewrite}
{\textrm tr}(dW^\dag\ dW) &=&
 {\ep^2\over 4}\,\cosh(\tau)\left[d\tau^2 + (\tilde{g}^5)^2
 + 2\left((\tilde{g}^3)^2+(\tilde{g}^4)^2\right)\right]  \\
 &+& {\ep^2\over 2}\,\sinh^2({\tau\over 2})\;
 \left[\; \left((\tilde{g}^1)^2+(\tilde{g}^2)^2\right)
 - \left((\tilde{g}^3)^2+(\tilde{g}^4)^2\right)\right]
 \nonumber\\
 |{\textrm tr}(W^\dag\ dW)|^2 &=&
 {\ep^4\over 4}\,\sinh^2(\tau) \left[\;d\tau^2\; +
 (\tilde{g}^5)^2 \right]
\eea
If we now use the fact that
\be
{1\over 3 K^2(\tau)} = {1\over 2} \cosh(\tau)\,K(\tau) + {1\over
2} \sinh(\tau)\,K'(\tau)
\ee
and plug back into (\ref{Kmetric}), we recover the metric
(\ref{mtmetric})
\be
ds_6^2 = {1\over 2}\varepsilon^{4/3} K(\tau)  \Bigg[ {1\over 3
K^3(\tau)} (d\tau^2 + (g^5)^2)  +  \cosh^2 \left({\tau\over
2}\right) [(g^3)^2 + (g^4)^2]  + \sinh^2 \left({\tau\over
2}\right)  [(g^1)^2 + (g^2)^2] \Bigg].
\ee

\section{String theory Hamiltonian}
The equations of motion following from the worldsheet Lagrangian of
section 5 are
\bea
&&\eta^{\a\b}\pd_\a \pd_\b x^i=0, \nonumber \\
&&\eta^{\a\b}\pd_\a \pd_\b z-(m_0\p^+\a')^2({4a_1\over a_0}-{4\over 5})z
=0, \nonumber \\
&&\eta^{\a\b}\pd_\a \pd_\b u_1-(m_0\p^+\a')^2({4a_1\over a_0}-{3\over 5})u_1
-\sqrt{2}m_0 p^+ \a' \left({a_1\over a_0}\right)^{1/2}\pd_\s v_2 =0, \nonumber \\
&&\eta^{\a\b}\pd_\a \pd_\b v_2-(m_0\p^+\a')^2 v_2
+\sqrt{2}m_0 p^+ \a' \left( {a_1\over a_0}\right)^{1/2}\pd_\s u_1 =0, \nonumber \\
&&\eta^{\a\b}\pd_\a \pd_\b u_2-(m_0\p^+\a')^2({4a_1\over a_0}-{3\over 5})u_2
-\sqrt{2}m_0 p^+ \a'\left( {a_1\over a_0}\right)^{1/2}\pd_\s v_1 =0, \nonumber \\
&&\eta^{\a\b}\pd_\a \pd_\b v_1-(m_0\p^+\a')^2 v_1
+\sqrt{2}m_0 p^+ \a' \left({a_1\over a_0}\right)^{1/2}\pd_\s u_2 =0,
\eea
The Fourier expansion for
generic closed-string classical solutions $z$ can be written as
\bea
z(\s,\tau ) &=& i \sqrt{\a' \over 2} \bigg[ {1 \ov \sqrt{\hat
m_z}} \big( a_0^z e^{-2i \hat m_z\tau } -
a_0^{i\dagger} e^{2i \hat m_z \tau}\big) \nonumber \\
&+& \sum_{n=1}^\infty {1\over
\sqrt{w_n^z}}\big[ e^{-2i w_n^z\tau }( a_n^i e^{2in\s } + \tilde
a_n^i e^{-2in\s }) - \ e^{2i w_n^z\tau }\big(
a_n^{i\dagger } e^{-2in\s } + \tilde a_n^{i\dagger} e^{2in\s }
\big)\big] \bigg] \ ,
\eea
where $w_n^z=\sqrt{n^2+ \hat m_z^2 }$. We have conveniently
normalized the $a$'s such that their Poisson bracket is not
proportional to the frequency. For the massless coordinates we
have the standard expansion. To determined the frequencies of the
coupled system we introduce the standard mode expansion
\be
u_1=\sum\limits_n A^1_n e^{i(\o_n\tau+ n\s)}, \quad
v_1=\sum\limits_n B^1_n e^{i(\o_n\tau+ n\s)},
\ee
and a similar ansatz for $u_2$ and $v_2$. Substituting the ansatze in
the equations of motions give
\be
(\omega_n^{\pm})^2 ={1\over 2}\bigg[2\,n^2+\hat m_v^2 +\hat m_u^2
\pm \sqrt{(\hat m_v^2-\hat m_u^2)^2+4\,n^2\,\hat m_B^2} \bigg],
\ee
where
\be
\hat m_v=m_0 p^+\a', \quad \hat m_u^2 =(m_0 p^+ \a')^2 ({4a_1\over
a_0}-{3\over 5}), \quad \hat m_B=\sqrt{2}m_0 p^+ \a'({a_1\over
a_0})^{1/2}.
\ee
Note that the frequency of the zero modes are $\omega_0^+=\hat
m_v$ and $\omega_0^-= \hat m_u$ and they correspond at the
zero-mode level to excitations associated with $v$'s and $u$'s
respectively. We therefore have three massless oscillators, one
massive with mass $\hat m_z$ , two massive with $\hat m_v$ and the
final two with $\hat m_u$.  As expected the effect of the B-field
($\hat m_B$) appears only for the nonzero modes ($n\ne 0$).

   For the fermionic sector (for more detail, see for example \cite{russo,cvj}) we have
two coupled Majorana-Weyl spinors $\t^1$ and $\t^2$ which satisfy
a the lightcone gauge condition, $\Gamma_+ \t^{1,2} = 0$, and
obey the following equation of motion:
\bea
(\pd_\tau + \pd_\sigma)\t^1 &=&
-{1\over 8}(p^+\alpha')\Gamma^{ij}(F_3)_{+ij}\,\t^2
-{1\over 8}(p^+\alpha')\Gamma^{ij}(H_3)_{+ij}\,\t^1, \\
(\pd_\tau - \pd_\sigma)\t^2 &=&
-{1\over 8}(p^+\alpha')\Gamma^{ij}(F_3)_{+ij}\,\t^1
+{1\over 8}(p^+\alpha')\Gamma^{ij}(H_3)_{+ij}\,\t^2 \nonumber.
\eea
In order to solve these equations, we combine the two spinors
into one complex spinor $\epsilon$, fourier transform $\epsilon$
in the variable $\sigma$ and combine the two first order equations
into one second-order equation.  For the MN plane--wave we get
\be
\ddot \epsilon_n = - \left[n^2 + {m_0^2\over 18}(5 - 3\Gamma_{u_1u_2v_1v_2})\right]\epsilon.
\ee
which gives four fermionic oscillators each with $\Gamma_{u_1u_2v_1v_2}=\pm 1$ and frequencies
\be
\omega^+_n = \sqrt{n^2 + {\hat m_0^2\over 9}},\qquad \textrm{and}\qquad
\omega^-_n = \sqrt{n^2 + {4\,\hat m_0^2\over 9}}
\ee
with $\hat m_0 = p^+\alpha'\,m_0$.  For the KS plane--wave we get the a slightly more complicated equation:
\be
\ddot \epsilon_n = - \Bigg[n^2 + \hat m_f^2 +\left\{-{1\over 4}\hat m_f^2 - n {\hat m_f \over 2}(i\Gamma_{u_1v_2}) - {\hat m_f^2\over 2}
(i\Gamma_{u_2v_2})\right\}(1 + \Gamma_{u_1u_2v_1v_2})\Bigg]\,\epsilon
\ee
with $\hat m_f = (2a_1/a_0)^{1\over2}\,m_0 p^+\alpha'$.  If we expand our spinors in a $\pm 1$ eigenbasis of $(i\Gamma_{u_1u_2})$ and \
$(i\Gamma_{u_1u_2})$ we can write down four pairs of fermionic oscillators.  The frequencies for the first two pairs have the simple
form
\be
\omega^{++}_n = \omega^{--}_n = \sqrt{n^2 + \hat m_f^2}.
\ee
while the spinors with eigenvalues $(+-)$ and $(-+)$ mix.  To get
their eigen--frequencies we diagonalize the matrix
\be
\left[
\begin{array}{cc}
(n^2 + {1\over 2} \hat m_f^2) & -(in\,\hat m_f + {1\over 2} \hat m_f^2) \\
(in\,\hat m_f - {1\over 2} \hat m_f^2) & (n^2 + {1\over 2} \hat m_f^2)
\end{array}
\right]
\ee
which yields the frequencies
\be
\omega^{"\pm"}_n = \sqrt{n^2 + {1\over 2} \hat m_f^2 \pm
{1\over 2}{\hat m_f}\sqrt{\hat m_f^2 + 4n^2}}.
\ee
Note that $\omega_0^{``-"}$ is a zero-frequency mode .

\section{Stability of the Wilson loop ansatz}

Let us now turn to the question of the reliability of the
results on the properties of the Wilson loop as a function of
$L$ and $J$. The first important issue we  clarify is the stability of
the classical configuration. This is most easily done in the
Polyakov formulation of 
the string. We thus consider the nonlinear sigma
model
\be
S={1\over 4\pi\a'}\int d^2\s\, G_{ij} \pd_a X^i \pd^a X^j.
\ee
In the conformal gauge, the equation of motion for $X^a$ being $t$, $\phi$ or $x$
has the same form:
\be
\pd_{\a}\left(g_{aa}\,\eta^{\a\b}\,\pd_\b\, X^a\right)=0.
\ee
They are all trivially satisfied. To understand the stability of the
solution we consider the linear fluctuations from the equations of
motion. Since $\pd_r g_{aa}|-{r_0}=0 $ we see that there is no mixing
between the
radial fluctuation and the fluctuations along $\phi$ and $x$. Next we
note that the resulting equation for the fluctuations ($\phi=\phi_0+
e^{i\nu\tau}\delta\phi(\s))$ is simply
\be
\label{fluc}
(\pd_\s^2 +\nu^2)\delta\phi(\s)=0.
\ee
We should now assume some boundary conditions for $\delta\phi(\s)$. The natural ones would be
Dirichlet at the ends of the interval where our solution is reliable, i.e. in
$\s=[\delta/\ell,2\pi -\delta/\ell]$. This implies that the spatial structure of the fluctuation is
\be
\delta\phi(\s)=\delta\phi_n\sin\bigg[{n\over 1 -{\delta\over \pi\ell}}(\s-{\delta\over \ell})\bigg].
\ee
Returning to (\ref{fluc}) we see that all the frequencies are positive and therefore the
solution is stable.
The last important point is  understanding the contribution of the edges, that is, the
regions in which it is not appropriate to consider $\pd_\s r\approx 0$
since precisely in these regions the strings go between the
boundary and
the ``minimal'' radius. The effects of these edges have been considered
in a similar setting in, for example, \cite{flat}. The estimate of
\cite{flat} applied to our situation gives $
{d\over L}\sim {\cal O}(L^{-1}),$
where $d$ is the region along which $\pd_\s r$ cannot be approximated
as zero. This means that in the large $L$ limit the contribution of
these regions is correspondingly suppressed. Similarly we expect suppression of the order of $1/J$.

\end{document}